\documentclass[aps,prl,twocolumn,superscriptaddress]{revtex4-2}
\usepackage{graphicx,amssymb,amsmath,amsfonts,empheq}
\usepackage[dvipsnames]{xcolor}
\usepackage{color}
\usepackage{braket}
\usepackage{latexsym}
\usepackage{upgreek}
\usepackage{dsfont}
\usepackage{bm}
\usepackage{makecell}
\usepackage{multirow}
\usepackage{enumitem}
\usepackage[normalem]{ulem}
\usepackage{url}
\usepackage{bbold}
\usepackage{hyperref}
\hypersetup{
colorlinks = true,
linkcolor = [rgb]{0,0,0}, 
citecolor = [rgb]{0.13,0.55,0.13},
urlcolor = [rgb]{0.25, 0.41, 0.88}}
\usepackage{bbm}

\usepackage{tikz}
\usepackage{tikz-cd}
\usetikzlibrary{arrows}
\usetikzlibrary{intersections}
\usetikzlibrary{shapes.geometric}
\usetikzlibrary{decorations.pathmorphing, patterns,shapes}
\usetikzlibrary{decorations.markings}

\tikzset{
	partial ellipse/.style args={#1:#2:#3}{
		insert path={+ (#1:#3) arc (#1:#2:#3)}
	}
}

\tikzset{
	mid arrow/.style={postaction={decorate,decoration={
				markings,
				mark=at position .575 with {\arrow[#1]{stealth}}
	}}},
	near arrow/.style={postaction={decorate,decoration={
				markings,
				mark=at position .275 with {\arrow[#1]{stealth}}
	}}},
	far arrow/.style={postaction={decorate,decoration={
				markings,
				mark=at position .800 with {\arrow[#1]{stealth}}
	}}},
}

\pgfdeclarepatternformonly{south west lines}{\pgfqpoint{-0pt}{-0pt}}{\pgfqpoint{3pt}{3pt}}{\pgfqpoint{3pt}{3pt}}{
	\pgfsetlinewidth{0.4pt}
	\pgfpathmoveto{\pgfqpoint{0pt}{0pt}}
	\pgfpathlineto{\pgfqpoint{3pt}{3pt}}
	\pgfpathmoveto{\pgfqpoint{2.8pt}{-.2pt}}
	\pgfpathlineto{\pgfqpoint{3.2pt}{.2pt}}
	\pgfpathmoveto{\pgfqpoint{-.2pt}{2.8pt}}
	\pgfpathlineto{\pgfqpoint{.2pt}{3.2pt}}
	\pgfusepath{stroke}}

\newcommand{\bbrakket}[2]{\mbox{$ \langle\!\langle #1 | #2 \rangle\!\rangle $}}

\newcommand{\kket}[1]{\mbox{$| #1 \rangle\!\rangle$}}
\newcommand{\bbra}[1]{\mbox{$\langle\!\langle #1 |$}}
\renewcommand{\bar}{\overline}

\renewcommand{\geq}{\geqslant}

\newcommand{\ri}{\mathrm{i}}

\newcommand{\tr}{\operatorname{tr}}

\newcommand{\bbZ}{\mathbb{Z}}

\newcommand{\calC}{\mathcal{C}}

\newcommand{\calG}{\mathcal{G}}

\newcommand{\calK}{\mathcal{K}}
\newcommand{\calL}{\mathcal{L}}
\newcommand{\calM}{\mathcal{M}}
\newcommand{\calN}{\mathcal{N}}

\newcommand{\calP}{\mathcal{P}}

\newcommand{\figref}[1]{Fig.~\ref{#1}}

\begin{document}

\title{Mixed-state topological order and the errorfield double formulation of decoherence-induced transitions}

\author{Yimu Bao}
\affiliation{Department of Physics, University of California, Berkeley, CA 94720, USA}
\author{Ruihua Fan}
\affiliation{Department of Physics, Harvard University, Cambridge, MA 02138, USA}
\author{Ashvin Vishwanath}
\affiliation{Department of Physics, Harvard University, Cambridge, MA 02138, USA}
\author{Ehud Altman}
\affiliation{Department of Physics, University of California, Berkeley, CA 94720, USA}
\affiliation{Materials Sciences Division, Lawrence Berkeley National Laboratory, Berkeley, CA 94720, USA}

\begin{abstract}
We develop an effective field theory characterizing the impact of decoherence on states with Abelian topological order and on their capacity to protect quantum information. The decoherence appears as a temporal defect in the double topological quantum field theory that describes the pure density matrix of the uncorrupted state, and it drives a boundary phase transition involving anyon condensation at a critical coupling strength. The ensuing decoherence-induced phases and the loss of quantum information are classified by the Lagrangian subgroups of the double topological order. Our framework generalizes the error recovery transitions, previously derived for certain stabilizer codes, to generic topologically ordered states and shows that they stem from phase transitions in the intrinsic topological order characterizing the mixed state.
\end{abstract}

\maketitle

Recent experiments in many-body quantum simulation platforms, such as arrays of Rydberg atoms and superconducting qubit systems, have shown compelling evidence for the establishment of topological order~\cite{Semeghini:2021wls,Bluvstein:2021jsq,Satzinger:2021eqy,GoogleQuantumAI:2022fyn}. 
While these platforms offer a high degree of control, the state preparation is subject to inevitable decoherence, leading ultimately to a mixed state. 
Experimental probes of topological order have a long history in the context of the fractional quantum Hall effect in semiconductors~\cite{Tsui:1982yy,AnyonCollisions,AnyonBraid}. 
In this case, it is understood that the imperfection due to non-vanishing temperature broadens the signatures of topological order~\cite{xiaogangReview,Nussinov2008PRB,Nussinov2009PNAS,Bravyi2009NoGo,Poulin2013PRL,Brown2016RMP}.
However, the states prepared in these new platforms do not have time to reach thermal equilibrium before they are probed and therefore cannot be described as Gibbs states. 
Instead, they are better modeled as a topologically ordered pure state corrupted by local channels describing decoherence for a finite time.
It is natural to ask how to characterize the topological order in such corrupted topological states.

Different perspectives on this problem suggest seemingly conflicting answers.
On one hand, the operation of the local quantum channels has an equivalent description as a finite-depth unitary process in an extended Hilbert space that includes an ancilla qubit for every system qubit. 
This finite-depth unitary circuit cannot lead to a singular change in the expectation value of any conventional diagnostic used to probe topological order, such as Wilson loops or open string operators~\cite{Fradkin:2013sab}, which support in a finite region. 
From this perspective, the topological order persists for any strength of decoherence, short of full dephasing or depolarization.

On the other hand, we can assess the effect of decoherence on the topological order through its ability to destroy the protection of quantum information encoded in the degenerate ground-state subspace~\cite{Kitaev:1997wr,Bravyi:1998sy}. 
In this vein, Dennis et al.~\cite{Dennis:2001nw} calculate a finite error threshold for the Toric code subject to local Pauli $X$ and $Z$ errors.
These errors create pairs of syndromes (i.e. anyon excitations) detected by stabilizer measurements. 
The recovery scheme is an algorithm for annihilating the syndromes in pairs, which fails when the error rate exceeds the threshold.
The existence of a recovery threshold strongly suggests that there is an underlying decoherence-induced transition in the intrinsic properties of the mixed state, which is not captured by standard probes used to detect ground-state topological order.
We note, however, that the recovery fidelity itself is not an intrinsic property, as it may depend on the recovery algorithm.
Furthermore, theories of recovery thresholds are limited to solvable topological stabilizer codes and do not offer insight into the effect of decoherence on topologically ordered states more generally.

In this Letter, we develop a universal description of topological order and decoherence-induced transitions in corrupted mixed states that is based on the underlying topological quantum field theory (TQFT). 
A key step is treating the density matrix as a state vector in a double (ket and bra) Hilbert space, $\kket{\rho}=\mathcal{N}\,|\Psi_0\rangle\otimes|\Psi_0^*\rangle$. 
The action of the decoherence channel $\mathcal{N}$, which couples the ket and bra states, can induce anyon condensation in the double space.
We argue that this transition is described in terms of boundary criticality in close analogy to the effect of measurement on quantum ground states analyzed in Ref.~\cite{Garratt:2022ycp}.

To derive an effective theory of the decoherence-induced transition, we utilize the description of the state $|\Psi_0\rangle$ as a (2+1)-dimensional TQFT. 
Then, $\kket{\rho}$ is given as a double TQFT coupled by the quantum channel $\mathcal{N}$ only at the temporal boundary $\tau=0$, and the induced transition  corresponds to anyon condensation on the boundary. 
Accordingly, the distinct decoherence-induced phases are classified by the possible boundary anyon condensates~\cite{Levin:2013gaa,Barkeshli:2013,Wang:2012am}. 
We note that the description in terms of boundary phase transitions explains why the error threshold transitions in stabilizer models invariably map to transitions in 2D classical statistical mechanics or (1+1)D quantum models, despite representing transitions in 2D quantum states~\cite{Dennis:2001nw,Wang:2002ph,Katzgraber:2009zz,Bombin:2012jk,Kubica:2018rab,Chubb:2021htd,CastelnovoPRB2008,Venn:2022kxy,Behrends:2022feh}. 

\begin{figure}[t]
\centering
\includegraphics[width=0.48\textwidth]{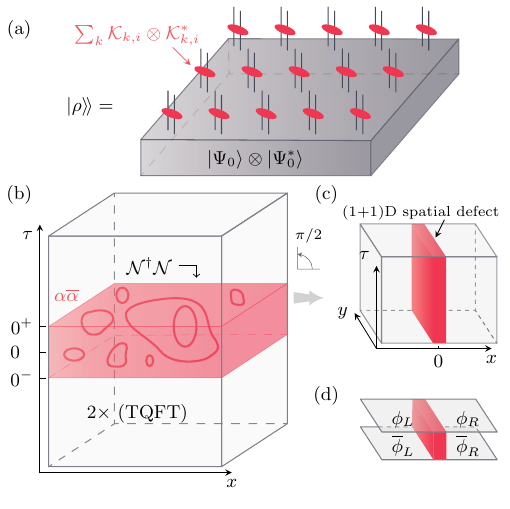}
\caption{(a) Errorfield double state $\kket{\rho}$ describing a topologically ordered pure state corrupted by local errors (red dots). (b) Path integral representation of the norm $\bbrakket{\rho}{\rho}$. Two path integrals in the past ($\tau < 0$) and the future ($\tau > 0$) prepare the double ground states at $\tau = 0^-$ and $0^+$. The states are coupled by the quantum channel $\calN^{\dagger}\calN$ (red cuboid). Red lines represent the worldlines of an anyon $\alpha\bar{\alpha}$ created by the incoherent errors. (c) The $\pi/2$ rotation converts the 2D temporal defect at $\tau = 0$ to a (1+1)D spatial defect at $x = 0$. (d) The rotated path integral represents the norm of the ground state of double 2D topological order with a 1D defect.}
\label{fig:efd}
\end{figure}

\emph{Errorfield double state.}--- 
The basic subject for our analysis is a topologically ordered state $\ket{\Psi_0}$ affected by local decoherence channels $\mathcal{N}_i$:
\begin{equation}
    \rho = \prod_i \calN_{i} [\ket{\Psi_0}\bra{\Psi_0}]\,,
\end{equation}
where $i$ is a site index. 
The local channel $\calN_i[\cdot] := \sum_k \calK_{k,i} (\cdot) \calK_{k,i}^\dagger$ is defined by the Kraus operators $\calK_{k,i}$, which satisfy $\sum_k \calK_{k,i}^\dagger \calK_{k,i} = \mathbb{1}$; they create different kinds of errors in the state.
For example, the channel that causes bit-flip errors consists of two Kraus operators $\calK_{1,i} = \sqrt{1-p}\,\mathbb{1}$ and $\calK_{2,i} = \sqrt{p}X_i$, where $p$ is the error rate. 

To understand the effect of decoherence on a quantum ground state, it is convenient to express the Kraus operators in terms of the quasiparticles they create. 
In particular, a Kraus operator acting on a topological state generally creates a superposition of different clusters of anyons, i.e. $\mathcal{K}_{k,i}=\sum_\alpha c_{k,\alpha}\, \mathcal{A}_i[\alpha]$, where $c_{k,\alpha}$ is a complex coefficient, and $\mathcal{A}_i[\alpha]$ creates a cluster of anyons in the vicinity of site $i$~\footnote{Strictly speaking, such a decomposition is well-defined for fixed-point states where the sizes of anyons are zero. In more general situations, e.g., Laughlin states, one can regard this decomposition as a \emph{definition} of phenomenological error models that we choose to work with.}.
Here, we focus on Kraus operators that are dominated by a single term rather than a coherent superposition, i.e. $\mathcal{K}_{k,i}$ is approximately proportional to a certain $\mathcal{A}_i[\alpha]$.
Such Kraus operators are said to create incoherent errors~\footnote{For example, one type of incoherent error creates a single anyon pair, an anyon and its anti-particle, on neighboring sites.}. 
The resulting mixed state is an incoherent mixture of states with different anyon configurations.

An essential step in characterizing the phases of the corrupted state is to regard the density matrix as a state vector in a double Hilbert space, i.e.
\begin{align}
	\kket{\rho} = \mathcal{N} \ket{\Psi_0} \otimes \ket{\Psi_0^*}
 = \prod_i \sum_{k} \calK_{k,i} \bar{\calK}_{k,i}\ket{\Psi_0} \otimes \ket{\Psi_0^*},
 \label{eq:EFD definition}
\end{align}
where we define barred operators as acting on the bra space, e.g. a super-operator $\calK (\cdot)\calK^\dagger$ acting on a density matrix is denoted as $\calK\,\bar{\calK}$ when acting on $\kket{\rho}$.
We refer to $\kket{\rho}$ as an \emph{errorfield double} (EFD) because of its analogy to the thermofield double state~\cite{Haag:1992hx}.
 
In the absence of decoherence, $\kket{\rho}$ contains two decoupled copies of conjugate topologically ordered states. 
The non-unitary evolution affected by the operator $\mathcal{N}$ can induce phase transitions at finite error rates that can be diagnosed by operator expectation values in the EFD. 
Up to normalization, these expectation values are of the form $\bbra{\rho}O\,\bar{O} \kket{\rho} = \tr(\rho O\rho O^\dagger)$.
This object is the overlap between $\rho$ and $\rho'= O\rho O^\dagger$, which measures the distinguishability between these two states.

\emph{Decoherence-induced transition as anyon condensation.}--- 
To begin, we consider the example of the Toric code~\cite{Kitaev:1997wr} subject to bit-flip errors.
The Kraus operator $X_i=m_r m_{r'}$ is incoherent and creates two $m$ anyons at neighboring locations $r,r'$. 
In the EFD, the corresponding super-operator creates two neighboring anyon pairs, $X_i\bar{X}_i=(m\bar{m})_r (m\bar{m})_{r'}$, where each pair $m\bar{m}$ creates an anyon of the double Toric code. 
The resulting EFD is given by
\begin{equation}
\kket{\rho} = \exp\left(\mu \sum_{\langle r r'\rangle}(m\bar{m})_r (m\bar{m})_{r'}\right) \ket{\Psi_0}\otimes \ket{\Psi_0^*},
\end{equation}
where $\mu = \tanh^{-1}[p/(1-p)]$. 
The operation of the error channel generates an effective imaginary time evolution of the pure topological state.
As shown in the supplementary material~\cite{SOM}, this leads to anyon condensation of $m\overline{m}$ at $p_c^{(2)} = 0.178$, with the solvable transition in the 2D (classical) Ising universality class~\footnote{The superscript of $p_c^{(2)}$ indicates the transition in EFD is detected by the quadratic function of the density matrix. For the general $n$-th moment of the density matrix, the critical point $p_c^{(n)}$ depends on $n$. It approaches the error-correction threshold $p_c = 0.109$ in the replica limit $n \to 1$~\cite{Diagnosticstoappear}.}~\footnote{The existence of a transition in the decohered mixed state at a finite error rate is in stark contrast with the destruction of 2D topological order in the Gibbs state at any finite temperature. In the double state $\kket{\rho}$, a far-separated pair of anyons is created with an exponentially decaying probability, while such an anyon pair occurs with a finite probability in the Gibbs ensemble regardless of the distance. The suppressed probability for separated anyons prevents logical errors and is the reason for the robustness of topological order against local decoherence.}. 

The anyon condensation can be detected by the open string operator
\begin{equation}
    \langle\!\langle W_{m\overline{m}}(\calP)\rangle\!\rangle := \frac{\bbra{\rho} W_{m\overline{m}}(\calP) \kket{\rho}}{\bbrakket{\rho}{\rho}} = \frac{\tr\rho\, w_m(\calP)\, \rho\, w_m^\dagger(\calP)}{\tr \rho^2},
    \label{eq:OS}
\end{equation}
where $w_{m}(\calP) := \prod_{i \in \calP} X_i$, and $W_{m\bar{m}} = w_m \otimes w_m^*$ creates the paired anyon $m\overline{m}$ at each end of the open path $\calP$~\footnote{We remark that the condensate can also be generally probed by the Fredenhagen-Marcu (FM) order parameter of $m\bar{m}$~\cite{Fredenhagen:1983sn,Fredenhagen:1985ft,Fredenhagen:1987gi,Gregor:2010ym,Verresen:2020dmk}.}.
In the topological phase, this quantity decays exponentially with the distance between two endpoints. 
In other words, the density matrix obtained by injecting an extra pair of far-separated anyons into $\rho$ is orthogonal to $\rho$ and hence distinguishable from it. 
In this sense, the $m$ anyons remain well-defined excitations.  
Once the paired anyon condenses, the open string saturates to a non-vanishing value at a long distance. 

Other types of incoherent errors in the Toric code, e.g. Pauli-Z or multi-qubit Pauli errors, can induce distinct phases with condensed electric anyon pairs $e\bar{e}$ or fermion pairs $f\bar{f}$.
All of these transitions are mutually independent because the paired anyons are mutual bosons, and they occur at the same $p_c^{(2)}$.
If two types of paired anyons condense simultaneously, it implies a condensate of the third type as well because each two are equivalent (i.e. fuse) to the third type, e.g. $e\times m=f$.

We remark that the anyon condensation can be also probed by the topological entanglement entropy (TEE) of the EFD~\cite{Kitaev:2005dm,LevinWen2005}, which changes discontinuously at the transition.
In the Toric code without decoherence, the TEE is $2\log 2$, with each of the two states contributing $\log 2$.
The condensation of $m\bar{m}$ reduces the TEE to $\log 2$, and further condensing also $e\bar{e}$ completely removes the TEE~\cite{SOM}.

The results in the above example can be readily extended to general Abelian topological orders subject to incoherent errors.
The paired anyon $\alpha\bar{\alpha}$ in the EFD are self-bosons and may condense at a large error rate.
Moreover, different types of paired anyons, $\alpha\bar{\alpha}$ and $\beta\bar{\beta}$, are always mutual bosons and can condense independently.

Anyon condensation in the EFD affects the information encoding in the corrupted state.
In topological quantum memory, the information is manipulated by the logical operator $w_\alpha(\ell)$ that moves an anyon $\alpha$ along non-contractible loops $\ell = \ell_1,\ell_2$ on the torus.
These logical operators are non-commuting and therefore encode quantum information.

In the condensed phase of $\alpha\bar{\alpha}$, $W_{\alpha\bar{\alpha}}$ acting on the EFD does not produce an orthogonal state, i.e. the two states have a finite overlap, indicating the loss of information encoded in $w_\alpha$.
Furthermore, the condensation identifies $W_{\beta\bar{\beta}}$ with $W_{\eta\bar{\eta}}$ provided that $\alpha \times \beta = \eta$.
Thus, two logical operators $w_\beta(\ell)$ and $w_\eta(\ell)$ are no longer independent.
By analyzing the commutation relations between independent logical operators, one can determine the encoding ability.
If the remaining operators are non-commuting, the corrupted state remains a quantum memory, otherwise, only classical information can be encoded. 
The memory is completely destroyed if no operators remain.

\emph{Effective field theory.---} To generalize the decoherence-induced transitions to other topologically ordered states and gain insight into the nature of the ensuing phases, we formulate the problem within an effective field theory. 
To this end, we employ the path integral representation of the normalization    
\begin{equation}
\bbrakket{\rho}{\rho} = \bbra{\rho_0} \, \calN^\dagger \calN \, \kket{\rho_0}\, ,
\end{equation}
which plays the role of a generating function for expectation values in the EFD. 
Since $\kket{\rho_0} = \ket{\Psi_0} \otimes \ket{\Psi_0^*}$ is a double topologically ordered state, it is given as a double TQFT in the Euclidean half spacetime $\tau<0$.
Similarly, $\bbra{\rho_0}$ is represented by a double TQFT in $\tau>0$. 
Therefore, in the absence of decoherence, the norm $\bbrakket{\rho_0}{\rho_0}$ is represented by two decoupled TQFT in the entire spacetime. 
The decoherence, represented by $\calN^\dagger \calN$, creates a temporal defect in the double TQFT, which couples the conjugate copies on the plane $\tau=0$ [\figref{fig:efd}(b)]. 
We note that the effect of measurements on a quantum ground state is similarly represented by a temporal defect in the Euclidean action~\cite{Garratt:2022ycp}.

The transition induced by the coupling at $\tau = 0$ must be a boundary transition.
This renders a distinction between the condensation of $\alpha\bar{\alpha}$ in the EFD and that in a quantum ground state.
The latter involves the proliferation of $\alpha\bar{\alpha}$ loops of arbitrary size in the $(2+1)$D spacetime, while the former is marked by the proliferation of small loops of the condensed object $\alpha\overline{\alpha}$, fluctuating near $\tau = 0$ [Fig.~\ref{fig:efd}(b)].
The distinction has consequences on the signatures of the decoherence-induced anyon condensate.  

In the Toric code example, the usual ground-state condensation of $m\bar{m}$ would lead to confinement of anyon $e$ that has non-trivial mutual statistics with $m\bar{m}$~\cite{Burnell2017Review}.
This would be detected, for example, by the Wilson loop $W_e(\calC) := \prod_{i\in \calC} Z_i \otimes \mathbb{1}$ along a contractible loop $\calC$. 
Once $m\bar{m}$ is condensed, $W_e(\calC)$ is affected by a finite density of large $m\bar{m}$ loops that pierce $\calC$, leading to exponential decay of $W_e(\calC)$ with an area law, i.e. $W_e(\calC)\sim e^{-b \text{Area}(\calC)}$.    
By contrast, the decoherence-induced condensate only involves the proliferation of small $m\bar{m}$ loops that only affect $W_e(\calC)$ along its perimeter leading to perimeter-law scaling, $W_e(\calC) \sim e^{-a|\calC|}$, regardless of the error rate.
Thus, the anyon condensation in the EFD does not lead to confinement, and the Wilson loop cannot probe the transition.

\emph{Mapping to (1+1)D boundary phases.---}
To classify the possible decoherence-induced phases it is convenient to perform a $\pi/2$-spacetime rotation of the (2+1)D TQFT, $\tau \to -x$ and $x \to \tau$. 
The bulk action is invariant under this rotation~\cite{Witten:1988TQFT,Fendley_2007, Shankar_2011}, which maps a problem with a temporal defect to a ground state with a spatial defect [\figref{fig:efd}(c)]. 
The incoherent errors translate to hermitian perturbations in the parent Hamiltonian of this ground state ~\cite{SOM}, localized on the defect line [Fig.~\ref{fig:efd}(d)].   
The decoherence-induced phases are therefore mapped to the quantum phases of a one-dimensional defect in a double topologically ordered state, which can be equivalently formulated as boundary phases of a quadrupled state~\cite{BeigiShor2011,Kapustin2011,SOM}.

\begin{table*}[t]
\centering
\begin{tabular}{c|c|c|c|c}
\hline
\hline
Model & Memory & \multicolumn{2}{c|}{Edge condensate (generators of Lagrangian subgroup)} & Error that realizes the phase\\
\hline
\multirow{5}{*}{\makecell{Toric code \\ $K_{\text{TC}} = \begin{pmatrix}
    0 & 2 \\
    2 & 0 \\
    \end{pmatrix}
  $}}  & Quantum & I & $e_Le_R,\, \bar{e}_L\bar{e}_R,\, m_L m_R,\, \bar{m}_L\bar{m}_R$ & No error\\
 & Classical  & II & $e_L\bar{e}_L,\, e_R\bar{e}_R,\, e_L\bar{e}_R,\, m_L\bar{m}_Lm_R\bar{m}_R$ & Incoherent $e$ error\\
 & Classical  & III & $m_L\bar{m}_L,\, m_R\bar{m}_R,\, m_L\bar{m}_R,\, e_L\bar{e}_Le_R\bar{e}_R$ & Incoherent $m$ error\\
 & Classical  & IV & $f_L\bar{f}_L,\, f_R\bar{f}_R,\, f_L\bar{f}_R,\, e_L\bar{e}_Le_R\bar{e}_R$ & Incoherent $f$ error\\
 & Trivial  & V & $e_L\bar{e}_L,\, e_R\bar{e}_R,\, m_L\bar{m}_L,\, m_R\bar{m}_R$ & Any two types of incoherent errors\\
\hline
\multirow{5}{*}{\makecell{Double semion\\ $K_{\text{DS}} = \begin{pmatrix}
    2 & 0 \\
    0 & -2 \\
    \end{pmatrix}
  $}} & Quantum & I & $m_{aL}m_{aR},\, \bar{m}_{aL}\bar{m}_{aR},\, m_{bL}m_{bR},\, \bar{m}_{bL}\bar{m}_{bR}$ & No error\\
 & Quantum  & II & $m_{aL}\bar{m}_{aL},\, m_{aR}\bar{m}_{aR},\, m_{bL}m_{bR},\, \bar{m}_{bL}\bar{m}_{bR}$ & Incoherent $m_a$ error\\
 & Quantum  & III & $m_{bL}\bar{m}_{bL},\, m_{bR}\bar{m}_{bR},\, m_{aL}m_{aR},\, \bar{m}_{aL}\bar{m}_{aR}$ & Incoherent $m_b$ error\\
 & Classical  & IV & $b_L\bar{b}_L,\, b_R\bar{b}_R,\, b_Lb_R,\, m_{aL}\bar{m}_{aL}m_{aR}\bar{m}_{aR}$ & Incoherent $b$ error\\
 & Trivial  & V & $m_{aL}\bar{m}_{aL},\, m_{aR}\bar{m}_{aR},\, m_{bL}\bar{m}_{bL},\, m_{bR}\bar{m}_{bR}$ & Any two types of incoherent errors\\
\hline
\multirow{2}{*}{\makecell{$\nu = 1/3$ Laughlin state\\ $K_{1/3} = \begin{pmatrix}
    3 \\
    \end{pmatrix}
  $}} & Quantum & I & $\eta_L \eta_R^2, \bar{\eta}_L \bar{\eta}_R^2$ & No error \\
  & Trivial & II & $\eta_L\bar{\eta}_L, \eta_R\bar{\eta}_R$ & Incoherent error for quasiparticles \\
\hline
\hline
\end{tabular}
\caption{Decoherence-induced phases in the Toric code, double semion model, and $\nu = 1/3$ Laughlin state subject to incoherent errors. Distinct phases are labeled by Lagrangian subgroups. The Toric code model contains four superselection sectors, $\{1,e,m,f\}$, where the anyon $e$ and $m$ are self-bosons and mutual semions, and $e\times m = f$ is a fermion. The double-semion model also has four sectors $\{1, m_a, m_b, b\}$, where $m_a, m_b$ are semion and anti-semion, and their fusion $m_a \times m_b = b$ is a boson. The $\nu = 1/3$ Laughlin state has three sectors $\{1, \eta, \eta^2\}$, where $\eta$ and $\eta^2$ are the quasiparticle and quasihole, respectively. $\alpha_s$ and $\bar{\alpha}_s$ denote the anyon $\alpha$ in the ket and bra Hilbert space, while $s = L, R$ represent the left and right copy. Here, only the generators of the Lagrangian subgroup are listed. The fusion of any two generators is also in the subgroup and is condensed on the edge. In the last column, we comment on the error channel that realizes each phase.}
\label{tab:toric_code_and_double_semion_phases}
\end{table*}

The possible phases can be understood using the field theory description.
The Lagrangian characterizing the low-energy physics has three contributions
\begin{equation}
\calL = \calL_0 + \calL_{1} + \calL_{\calN}.
\end{equation}
$\calL_0$ describes the low-energy excitations in the TQFT, which involve only the edge degrees of freedom since the bulk is gapped.
$\calL_1$ represents the coupling between the edge modes on the left and right of the defect in Fig.~\ref{fig:efd}(d), which stems from the continuity of the TQFTs in the two half spacetimes before the rotation. 
This is the coupling ensuring that the system before decoherence is described by a gapped theory without temporal boundaries.
Lastly, $\calL_\calN$ represents the effect of the error channels, which couple the edge modes in the ket and bra copies.

We now focus on the case of Abelian topological order, which is captured by Abelian Chern-Simons theories. 
The low-energy excitations are described by a theory of compact bosons on the edge~\cite{Elitzur:1989nr,Wen:1995qn}
\begin{align}
\mathcal{L}_0[\phi] = \frac{1}{4\pi} \sum_{I,J} \mathbb{K}_{IJ}^{(2)} \ri\partial_\tau \phi^I \partial_y \phi^J - \mathbb{V}_{IJ}^{(2)} \partial_y \phi^I \partial_y \phi^J,
\label{eq:edge_lagrangian}
\end{align}
where $\phi := [\bar{\phi}_R,\phi_R,\phi_L,\bar{\phi}_L]$, $\phi_{s}$ and $\bar{\phi}_{s}$ are the field variables in the ket and bra Hilbert spaces, respectively, while $s = L,\,R$ denote the two copies of the EFD.
A central object of the theory is the $K$-matrix $\mathbb{K}^{(2)} = K \oplus (-K) \oplus K \oplus (-K)$ with $K$ an integer-valued matrix.
$\mathbb{V}^{(2)} = V \oplus V \oplus V \oplus V$ with $V$ a non-universal positive definite matrix.
In this theory, $\calL_1$ is a non-linear coupling of $\phi_{L}$ to $\phi_R$ and $\bar{\phi}_L$ to $\bar{\phi}_R$, and $\mathcal{L}_\calN$ couples field variables within the same copy of the EFD, i.e. $\phi_s$ and $\bar{\phi}_s$.

The gapped phase of the defect is obtained by condensing bosonic excitations.
The excitations in $\calL_0$ are of the form $e^{\ri\bm{l}^T\cdot \phi}$ labeled by an integer vector $\bm{l}$.
Such excitations are bosonic if their self statistics $\theta_{\bm{l}} := \pi \bm{l}^T (\mathbb{K}^{(2)})^{-1}\bm{l} = 0 \mod 2\pi$.
For Abelian topological order, the condensed bosonic excitations on the edge form a group termed the Lagrangian subgroup~\cite{Levin:2013gaa,Barkeshli:2013,Wang:2012am}.

The possible edge phases are classified by inequivalent Lagrangian subgroups $\calM$.
In our case, the allowed subgroup is subject to symmetry constraints, i.e. it is invariant under the global symmetry $\calG^{(2)} = \mathbb{Z}_2 \times \mathbb{Z}_2^\mathbb{H}$ of the defect theory. 
Here, $\mathbb{Z}_2$ is an anti-unitary symmetry associated with the Hermitian conjugation in the double Hilbert space, i.e. $\phi_R^I(\bar{\phi}_R^I) \leftrightarrow \phi_L^I(\bar{\phi}_L^I)$ and $\ri \leftrightarrow -\ri$, and $\mathbb{Z}_2^{\mathbb{H}}$ is an anti-unitary symmetry originating from the hermiticity of density matrix and acts as $\phi_{s}^I \leftrightarrow \bar{\phi}_{s}^I$ with $s = L, R$ and $\ri \leftrightarrow -\ri$~\footnote{The K-matrix $\mathbb{K}^{(2)}$ flips sign under the anti-unitary symmetry, however, the Lagrangian $\calL_0$ is invariant as the minus sign cancels that from the imaginary identity.}~\footnote{The effective theory for the norm of the EFD $\bbrakket{\rho}{\rho} = \tr\rho^2$ is expected to have a $\mathbb{Z}_2$ unitary symmetry permuting two copies of density matrix. This permutation is a combination of the Hermitian conjugation of the EFD and that of the density matrix and thus is included in $\calG^{(2)}$.}.
Furthermore, the interaction only creates specific excitations: paired anyon $\alpha_s \overline{\alpha}_s = e^{\ri \bm{l}^T\cdot (\phi_s-\bar{\phi}_s)}$ by the incoherent errors, and $\alpha_L\alpha'_R = e^{\ri \bm{l}^T\cdot (\phi_L + \phi_R)}$, $\bar{\alpha}_L\bar{\alpha}'_R = e^{-\ri \bm{l}^T\cdot (\bar{\phi}_L +\bar{\phi}_R)}$ by $\calL_1$.
%
%
Hence, the edge phases are governed by the $\calM$ in which the condensed objects are the fusion of such paired anyons.

The Lagrangian subgroup $\calM$ must satisfy the following criteria:
\begin{enumerate}
	\item $e^{\ri \theta_{\bm{m}\bm{m'}}} = 1, \forall \bm{m}, \bm{m'} \in \calM$;
	\item $\forall \bm{l} \notin \calM$, $\exists \bm{m}$ s.t. $e^{\ri \theta_{\bm{ml}}} \neq 1$;
        \item $\forall \bm{m} \in \calM$, $ g\bm{m} g^{-1} \in \calM, \forall g \in \calG^{(2)}$;
	\item (Incoherent error) $[\bm{1},\bm{1},-\bm{1},-\bm{1}]^T \cdot \bm{m} = 0 \mod \mathbb{K}^{(2)}\Lambda, \forall \bm{m} \in \calM$~\footnote{The equation is held modulo physical excitations, where $\Lambda$ is an integer vector. $\bm{1}$ is a vector with each component being $1$ and of the same dimension as the matrix $K$.}.
\end{enumerate}
Here, $\theta_{\bm{m}\bm{l}} := 2\pi \bm{m}^T (\mathbb{K}^{(2)})^{-1} \bm{l}$ characterizes the mutual statistics between two anyons.
In the three examples, the Toric code, double semion model~\cite{LevinGu:2012}, and $\nu = 1/3$ Laughlin state~\cite{Laughlin1983}, we enumerate the possible phases (i.e. Lagrangian subgroups) summarized in Table~\ref{tab:toric_code_and_double_semion_phases}~\cite{SOM}.

\emph{Discussion.---} 
We have introduced the errorfield double formulation to characterize the breakdown of quantum memories described by Abelian topological orders, generalizing error recovery transitions in specific quantum codes. 
We showed that the breakdown of quantum memories corresponds to a boundary transition in a replicated theory consisting of $n\geq 2$ copies of the topological order and its conjugate. 
We focused on the $n=2$ case and incoherent errors and classified the possible phases.
Our results are ripe for multiple lines of generalization.

One useful extension is to characterize the effect of quantum channels that include coherent errors~\cite{Bravyi2018Coherent,Flammia2019QECCoherent,Preskill2020Coherent,Venn:2022kxy,Behrends:2022feh}. 
A physically important example is provided by amplitude damping errors applied to the Toric code~\cite{Poulin2017,Poulin2018}. 
Here, the decoherence involves terms that can drive condensation of anyons in a single-copy Hilbert space (i.e. $\alpha$ or $\bar{\alpha}$).
The corresponding Lagrangian subgroup does not need to satisfy the fourth criterion, allowing for the establishment of additional decoherence-induced phases~\cite{SOM}.

Another future direction is to investigate the effect of decoherence in non-Abelian topologically ordered states.
Here, decoherence can affect the capacity to perform protected quantum computation in the fusion space~\cite{Nayak:2008zza}.
One complication is that errors are intrinsically ``coherent" as incoherent errors can generate anyons in the individual ket or bra copy due to non-Abelian fusion rules.

The existence of an error recovery threshold suggests that the decoherence-induced phases can be characterized by information-theoretic properties. 
In a separate work, we propose information-theoretic diagnostics of the topological order in corrupted mixed states and demonstrate their consistency in a concrete example of the Toric code under incoherent errors~\cite{Diagnosticstoappear}. 
The EFD framework allows computing quadratic functions of the density matrix and therefore corresponds to the  second R\'enyi versions of the information-theoretic quantities.
However, the framework is generalizable to a replicated theory of the (1+1)D defect, allowing us to analyze the $n$-th moment of $\rho$ (see supplementary material~\cite{SOM}). 
The information-theoretic quantities are obtained in the limit $n\to 1$. 
We expect that the classification in Table~\ref{tab:toric_code_and_double_semion_phases} still holds in this limit~\cite{SOM}.

Finally, the EFD formalism can be applied to the investigation of mixed states beyond topologically ordered systems.
Decoherence-induced transitions may occur in other states that encode information nonlocally, such as fracton systems~\cite{Haah:2011drr,Vijay:2016phm} and quantum low-density parity check codes~\cite{Breuckmann:2021yvk}. 
The recently proposed average symmetry-protected topological phases~\cite{Ma:2022pvq,lee2022symmetry,zhang2022strange} is another problem that can be possibly characterized in the EFD formulation.

\begin{acknowledgments}
\emph{Acknowledgements}.---We thank Meng Cheng and Sam Garratt for helpful discussions. AV was funded by the Simons Collaboration on Ultra-Quantum Matter, which is a grant from the Simons Foundation (651440, AV). AV and RF further acknowledge support from NSF DMR-2220703.
Support is also acknowledged from the U.S. Department of Energy, Office of Science, National Quantum Information Science Research Centers, Quantum Systems Accelerator (EA).
EA and YB were supported in part by NSF QLCI program through grant number OMA-2016245.
This work is funded in part by a QuantEmX grant from ICAM and the Gordon and Betty Moore Foundation through Grant GBMF9616 to Yimu Bao and Ruihua Fan. 

\emph{Note added}.--- Upon completion of the first version of this manuscript, we became aware of an independent and related work~\cite{lee2023quantum}, which appeared as a preprint on the same day. 

Since this paper first appeared as a preprint, various generalizations have 
been formulated ~\cite{ellison2025toward,sohal2025noisy,wang2025intrinsic,sala2025decoherence,sala2025stability} as well as work studying complementary aspects of mixed state phases of matter~\cite{chen2024separability,chen2024symmetry,sang2025stability,wang2025analog,li2025replica,zhang2025strong,li2025much,lessa2025strong}.
\end{acknowledgments}

\bibliography{refs_short.bib}

\noindent\textbf{End Matter.} Since this paper first appeared as a preprint on arXiv, the study of mixed-state phases has become a very active field.
Here, we summarize the recent developments in this field and explain how these works relate to the current manuscript.

Mixed-state topological order formulated in this work is characterized by the quantities that are non-linear functions of the density matrix.
In particular, the transition in the double state is probed by R\'enyi-2 quantities that are quadratic functions of the density matrix.
These R\'enyi-2 quantities are often easy to study and help map out the landscape of the mixed-state phases. 
However, the R\'enyi-$n$ quantities (with $n \geq 2$) exhibit different thresholds and lack clear information-theoretical meanings compared to their von-Neumann counterparts (often realized in the limit R\'enyi index $n \to 1$).
One crucial theoretical question is to identify quantities with clear operational meanings in information-theoretical tasks to characterize mixed-state phases.
To address this question, in the companion paper~\cite{Diagnosticstoappear}, we proposed several quantities, including coherent information and topological entanglement negativity, as diagnostics of the mixed-state topological order.
Building on our result, later works further explore various mixed-state phases characterized by R\'enyi-$n$ quantities as well as information-theoretical quantities in the $n \to 1$ limit~\cite{wang2025intrinsic,sohal2025noisy,ellison2025toward,wang2025analog,zhang2025strong,li2025replica,sala2025decoherence,sala2025stability,wang2025fractional}.

The operational meaning of mixed-state topological order below the threshold and the decoherence-induced transition has been further investigated from several other perspectives.
%
%
First, Ref.~\cite{chen2024separability,chen2024symmetry} proposed that the intrinsic transition at the finite decoherence threshold $p_c$ (in the $n \to 1$ limit) can be formulated as a transition in the many-body separability.
Ref.~\cite{chen2024separability} showed that, above $p_c$, the decohered toric code becomes \emph{separable}; it exhibits a decomposition in terms of short-range entangled states (each of which can be prepared from a product state using a low-depth unitary circuit).
%
%
Second, Ref.~\cite{wang2025analog} proposed the convex-roof extension of quantum conditional mutual information as a probe of topologically ordered mixed states below the threshold, which generalizes the concept of topological entanglement entropy in topologically ordered pure states.
Furthermore, Ref.~\cite{sang2025stability} demonstrated that the decoherence-induced transition also manifests in the conditional mutual information (CMI) $I_{A:C|B}$ in the decohered toric code.
The CMI decays exponentially in the size of the ``buffer'' region $B$, i.e. $I_{A:C|B} \sim e^{-\ell/\xi}$, with a characteristic length scale $\xi$ called Markov length, both in the regime $p < p_c$ and $p > p_c$.
The Markov length $\xi$ diverges at the critical point $p_c$, demonstrating the diverging length scale at the decoherence-induced transition.
%

Having identified characteristics of the phases and phase transitions, a remaining question is to establish the equivalence relation between mixed states belonging to the same phase.
%
%
In quantum ground states, two states in the same phase are connected by low-depth unitary circuits.
Analogously, one can define the mixed-state phases by requiring the mixed states in the same phase to be connected by low-depth quantum channel circuits.
However, since quantum channels are in general irreversible, one needs a modified definition for mixed states.
Currently, two-way connection by quantum channels~\cite{coser2019classification,ma2023average,sang2024mixed,sang2025stability} and the connection by reversible quantum channel circuits~\cite{sang2025mixed} emerge as definitions of equivalent mixed states in the same phase; their subtle distinction is explained in Ref.~\cite{sang2025mixed}.

Our work also motivates the study of mixed-state topological order in various other setups.
Ref.~\cite{wang2025intrinsic} studied the toric code under fermion decoherence (Phase IV in Table~I of the main text) and showed that this phase, despite only encoding classical information, exhibits long-range quantum entanglement measured by topological entanglement negativity.
The existence of long-range entanglement in such a phase is later formulated rigorously, as a result of the strong anomalous 1-form symmetry generated by the fermion loop operators~\cite{li2025much}.
In particular, Ref.~\cite{li2025much} proved that the fidelity between the fermion decohered state and an ensemble of short-range entangled states decays exponentially in the system size.

The example of fermion decohered toric code demonstrates that the topological mixed states can exhibit 1-form symmetries (deconfined anyons) that form a premodular category (with non-trivial transparent anyons).
These mixed-state phases are named ``intrinsic mixed-state topological order''.
Based on this observation, Ref.~\cite{sohal2025noisy,ellison2025toward} proposed a classification of mixed-state topological order according to the premodular anyon theories.

The errorfield double formulation in this paper is also used to map out the landscape of decoherence-induced phases in other topologically ordered states.
Refs.~\cite{sala2025decoherence,sala2025stability} study non-Abelian topologically ordered states under local decoherence and identify the condition under which the topological order is retained in the mixed state up to the maximum decoherence strength.
Ref.~\cite{wang2025fractional} investigates various types of fractional quantum Hall states under density dephasing and maps out a phase diagram.
The distinct phases are naturally understood as boundary phases of the 2+1D path integral induced by local decoherence as proposed in the current manuscript.


The transition of mixed-state topological order in this work is also closely tied to the concept of strong-to-weak spontaneous symmetry breaking (strong-to-weak SSB) transition formulated in recent works~\cite{lee2023quantum,lessa2025strong}.
In Sec.~SI of the supplementary material, we study the phase transition in the toric code subject to Pauli-X decoherence channels in the doubled-state formulation. 
%
%
The condensation transition of $m\bar{m}$ at the finite threshold $p_c^{(2)}$ confines the anyon $e$ and $\bar{e}$ in the doubled state, which restores the 1-form symmetry generated by the loop operator of magnetic anyon $m\bar{m}$ (explicitly formulated in Ref.~\cite{zhang2025strong}).

This transition in the decohered toric code is dual to the strong-to-weak SSB transition in the decohered Ising paramagnet.
Specifically, the Kramers-Wannier dual of this problem concerns a product state $\ket{+}^{\otimes N}$ on the 2D square lattice subject to dephasing channels $\calN_{ZZ}[\cdot] = (1-p)[\cdot] + p Z_{i}Z_{j}[\cdot]Z_iZ_j$ acting on nearest-neighbor pairs $\langle i,j\rangle$.
The mixed state exhibits a strong $\bbZ_2$ global symmetry generated by $\prod_i X_i$.
The decoherence channel induces a transition in the doubled state at $p_c^{(2)}$ which breaks the strong $\bbZ_2$ symmetry down to the weak $\bbZ_2$ symmetry.
Refs.~\cite{lessa2025strong,weinstein2025efficient,liu2025diagnosing} further propose information-theoretical quantities in the limit R\'enyi index $n \to 1$, such as the fidelity correlator and correlators in the canonical purification, to diagnose the strong-to-weak SSB transition at $p_c$.

\end{document}


\title{Supplementary Online Material for ``Mixed-state topological order and the errorfield double formulation of decoherence-induced transitions''}
\date{\today}

\author{Yimu Bao}
\affiliation{Department of Physics, University of California, Berkeley, CA 94720, USA}
\author{Ruihua Fan}
\affiliation{Department of Physics, Harvard University, Cambridge, MA 02138, USA}
\author{Ashvin Vishwanath}
\affiliation{Department of Physics, Harvard University, Cambridge, MA 02138, USA}
\author{Ehud Altman}
\affiliation{Department of Physics, University of CA, Berkeley, CA 94720, USA}
\affiliation{Materials Sciences Division, Lawrence Berkeley National Laboratory, Berkeley, CA 94720, USA}

\maketitle

\tableofcontents

\section{Phase transition in Toric code subject to incoherent errors}
In this section, we provide details of the phase transition in the Toric code subject to bit-flip errors.
In Sec.~\ref{sec:transition}, we show that the anyon condensation transition in the EFD is exactly solvable.
It occurs at $p_c^{(2)} = 0.178$ and is of the 2D classical Ising universality.
In Sec.~\ref{sec:TEE}, we compute the topological entanglement entropy in the EFD state and show it sharply changes from $2\log 2$ to $\log 2$ at the condensation transition.

\subsection{Critical point}\label{sec:transition}
As mentioned in the main text, the bit-flip errors in the Toric code can induce a condensation of $m\bar{m}$-anyon in the EFD at a finite error threshold.
The transition is exactly solvable and of 2D classical Ising universality. 
Here, we provide details on exactly solving the transition.

To start, we consider the EFD state describing the Toric code subject to bit-flip (Pauli-X) errors
\begin{equation}
\label{eq:EFD_TC_bitflip}
    \kket{\rho} = \prod_{l} \left( (1-p) + p X_l\otimes X_l \right) \ket{\Psi_0}\otimes \ket{\Psi_0^*}\, \propto e^{\mu \sum_i X_l \otimes X_l} \ket{\Psi_0}\otimes \ket{\Psi_0^*},
\end{equation}
where $p$ is the bit-flip error rate, $\mu = \tanh^{-1}[p/(1-p)]$, $l$ labels the edges in the square lattice, and $\ket{\Psi_0}$ is one of the Toric code ground states specified below.

We work in a loop picture of the Toric code ground state
\begin{align}
\ket{\Psi_0} \propto \prod_p \frac{1 + B_p}{2} \ket{+} \propto \sum_{\{g\}} \ket{g},
\end{align}
where $p$ in the subscript labels the plaquette, $\ket{+} := \prod_i \ket{+}_i$ is a product state with each qubit initialized in the $+1$ eigenstate of Pauli-X operator, i.e. $X_i \ket{+} = \ket{+}$.
A product of $B_p$ gives $g_z$, which is a product of the Pauli-Z operators along a closed loop.
When acting on $\ket{+}$, $g_z$ flips qubits into $-1$ eigenstates of $X_i$ along a loop $g$ and leads to a product state $\ket{g} := g_z \ket{+}$. 
This allows writing the ground state in terms of closed loops $g$, represented by binary variables.

The EFD can be written in the loop picture accordingly.
Before applying the error channel, the double ground state is expressed as
\begin{align}
\kket{\rho_0} = \ket{\Psi_0} \otimes \ket{\Psi_0^*} = \frac{1}{\sqrt{\calZ(0)}}\sum_{g,\bar{g}} \kket{g, \bar{g}},
\end{align}
where $\kket{g, \bar{g}} := (g_z \otimes \bar{g}_z) \ket{+}\otimes \ket{+}$, $g$ and $\bar{g}$ label two independent loop configurations in the ket and bra copy, respectively.
The error channel assigns a weight in the summation of loops and leads to 
\begin{align}
    \kket{\rho} = \frac{1}{\sqrt {\calZ(\mu)}}\sum_{g,\bar{g}} e^{-2\mu|g+\bar{g}|}\kket{g,\bar g},
\end{align}
where $g+\bar{g}$ is a binary sum and represents the relative loop configuration, $\calZ(\mu)$ is the normalization factor, and $\mu$ can be interpreted as the loop tension.
We note that only the relative loop $h := g+\bar{g}$ acquires a loop tension.
Hence, it is more convenient to rewrite the EFD as
\begin{align}
    \kket{\rho} = \frac{1}{\sqrt{\calZ(\mu)}} \sum_{g,h} e^{-2\mu|h|}\kket{g,h}.
\end{align}

The anyon condensation transition is detected by the open string operator
\begin{align}
\langle\!\langle W_{m\overline{m}}(\calP)\rangle\!\rangle = \frac{\bbra{\rho} w_{m}(\calP) \otimes w_{\overline{m}}(\calP) \kket{\rho}}{\bbrakket{\rho}{\rho}} = \frac{1}{\calZ(\mu)}\sum_{g,h} (-1)^{\# \text{crossing}(\calP,h)} e^{-4\mu|h|},
\end{align}
where $\# \text{crossing}(\calP,h)$ counts the number of crossings between the open string $\calP$ and loop $h$.
In this expression, the open string $\langle\!\langle W_{m\overline{m}}(\calP)\rangle\!\rangle$ can be interpreted as an observable in a statistical mechanical model of loops.
The normalization factor $\calZ(\mu)$ is the partition function $\calZ(\mu) = \Omega(0)\Omega(\mu)$, where the partition function $\Omega(0)$ of $g$ loops is a constant independent of $\mu$, and the partition function of $h$ loop, $\Omega(\mu) = \sum_h e^{-H(\mu)}$, has an effective Hamiltonian $H(\mu) = 4\mu|h|$.

We can identify the loop configuration $h$ with domain walls in a spin model with Ising degrees of freedom living on the plaquettes, i.e.
\begin{align}
\begin{tikzpicture}[scale=0.75, baseline={(current bounding box.center)}]
\definecolor{myred}{RGB}{240,83,90};
\definecolor{myblue}{RGB}{93,123,189};
\small
\foreach \y in {0,0.7,1.4,2.1}{
	\draw[black!35] (-0.2,\y) -- (2.2,\y);
}
\foreach \x in {0,0.7,1.4,2.1}{
	\draw[black!35] (\x,-0.2) -- (\x,2.2);
}
\draw[myblue,line width=0.6] (-0.2,0.7) -- (0.7,0.7) --  (0.7,1.4) -- (2.1,1.4) -- (2.1,2.2);
\draw[myblue,line width=1.5] (0.7,1.4) -- (1.4,1.4);
\filldraw (1.05,1.05) circle (0.03) node[right]{\scriptsize$\sigma_i$};
\filldraw (1.05,1.8) circle (0.03) node[right]{\scriptsize$\sigma_j$};
\draw[dotted,line width=1] (1.05,1.05) -- (1.05,1.8);
\node[myblue] at (0.3,1.8) {$h_{l}$};
\end{tikzpicture}, \qquad h_l = \frac{1 - \sigma_i\sigma_j}{2}.
\end{align}
Then, the loop model reduces to the square lattice Ising model up to a constant
\begin{align}
H = \sum_{\langle i, j\rangle} -2\mu \sigma_i\sigma_j.
\end{align}
Thus, the condensation transition in the EFD is of the 2D classical Ising universality class.
The critical point is exactly solvable and determined by the Kramers-Wannier duality, $\mu_c = (1/4)\log(1+\sqrt{2})$, 
i.e. $p_c^{(2)} = 0.178$.
We remark that this critical point is associated with the partition function $\calZ(\mu) = \tr \rho^2$ and indicates the singularity in the second moment of $\rho$.

\subsection{Topological entanglement entropy}\label{sec:TEE}
Another diagnostic of the anyon condensation is the topological entanglement entropy (TEE) of the EFD state.
In the case of the Toric code subject to incoherent errors, the EFD state can be written as the ground state of a gapped local Hamiltonian.
One can therefore use TEE to detect the transition~\cite{Kitaev:2005dm,LevinWen2005}.
In this subsection, we first construct the parent Hamiltonian of the EFD.
We then compute the R\'enyi-$n$ entropy of the EFD and show that the subleading topological term undergoes a transition from $2\log 2$ to $\log 2$ upon the condensation of $m\bar{m}$ regardless of the R\'enyi order $n$.
This indicates a singularity also in the topological term of von Neumann entropy ($n\to 1$).
We remark that, in a related problem of the Toric code undergoing imaginary time evolution, the TEE changes sharply from $\log 2$ to zero across the phase transition as shown in Ref.~\cite{CastelnovoPRB2008}.

The EFD describing the Toric code subject to bit-flip errors is the ground state of a local parent Hamiltonian~\footnote{We thank Meng Cheng for pointing this out to us.}
\begin{align}
    \tilde{H} = \sum_{s} V^{-1} h_s V V h_s V^{-1} + \sum_p V^{-1} h_p V V h_p V^{-1},
\end{align}
where $V = e^{\mu \sum_i X_i \otimes X_i}$. 
The Hamiltonian term $h_s$ runs over $(1 - A_s\otimes \bbI)/2$ and $(1 - \bbI \otimes A_s)/2$, and $h_p$ runs over  $(1 - B_p\otimes \bbI)/2$ and $(1 - \bbI \otimes B_p)/2$.
Both $h_s$ and $h_p$ are projectors. 
In the case without decoherence, the EFD for the Toric code pure state is the zero-energy ground state of the Hamiltonian
\begin{align}
    H_0 = \sum_s h_s + \sum_p h_p = \sum_s \frac{1 - A_s\otimes \bbI}{2} + \frac{1 - \bbI \otimes A_s}{2} + \sum_p \frac{1 - B_p\otimes \bbI}{2} + \frac{1 - \bbI \otimes B_p}{2}.
\end{align}
Here, the eigenvalue of $\tilde{H}$ is non-negative. One can show $\kket{\rho}$ is a zero-energy eigenstate of $\tilde{H}$ and therefore is the ground state.
It is then legitimate to use the TEE to characterize the topological order in the EFD. 
The TEE sharply changes when the gap of $\tilde{H}$ closes.

We now compute the TEE in the EFD and show it undergoes a transition when tuning the error rate.
First, we show the TEE takes distinct values in two limits, the error rate $p = 0$ and $p = 1/2$.
To gain an understanding away from these two limits, we then consider the R\'enyi-$n$ entropy and map it to the free energy cost of spin pinning in a classical spin model.
We show its subleading term undergoes a transition at the critical point and take the analytic limit $n \to 1$ to extract the behavior of the TEE.
We remark that, as far as the entanglement entropy of a finite disk is concerned, the boundary condition of the total system is not important and will not be specified explicitly throughout the entire calculation.

In the limits $p = 0, 1/2$, the EFD state is a stabilizer state, and the TEE can be easily determined.
In the absence of errors, the EFD contains two copies of decoupled Toric code states and exhibits a TEE $\gamma = 2 \log 2$.
In the case of $p = 1/2$, the tension $\mu \to \infty$ imposes the stabilizers $X_i\otimes X_i$ in the EFD.
The resulting EFD is the mutual eigenstate of stabilizers
\begin{equation*}
\{A_s \otimes \mathbb{I}, \quad \mathbb{I} \otimes A_s, \quad X_i \otimes X_i, \quad B_p \otimes B_p\}.
\end{equation*}
For a simply connected region $A$, its entanglement entropy is given by
\begin{align}
    S_A/\log 2 = 2|A| - \# \text{stabilizer}(\text{supp}(s) \in A)
\end{align}
where $2|A|$ counts the total number of degrees of freedom in region $A$, and $\# \text{stabilizer}(\text{supp}(s) \in A)$ is the number of stabilizers whose support is completely in $A$.
One can show $S_A = (|\partial A|-1)\log 2$, and the TEE $\gamma = \log 2$.
The distinct values of TEE in two limits indicate its potential singularity as a function of error rate.

To understand the TEE away from two limits, we consider the R\'enyi-$n$ entropies of a simply connected region $A$ and develop a classical spin model description.
The first step is to write down the reduced density matrix.
Here, we separate the loop configuration $g$ ($h$) into two parts $g_A$ and $g_{\bar{A}}$ ($h_A$ and $h_{\bar{A}}$), which support on region $A$ and its complement $\bar{A}$, respectively.
To form a closed loop configuration, two parts must agree on the boundary of $A$, i.e., $\partial g_A = \partial g_{\bar A} = b$, and $\partial h_A = \partial h_{\bar A} = c$. 
\begin{equation*}
\begin{tikzpicture}[scale=0.45, baseline={(current bounding box.center)}]
\definecolor{myred}{RGB}{240,83,90};
\definecolor{myblue}{RGB}{93,123,189};
\definecolor{mygreen}{RGB}{83,195,189};
\small
\foreach \y in {0.7,1.4,...,4.9}{
	\draw[black!40] (0.7-0.2,\y) -- (5.1,\y);
}
\foreach \x in {0.7,1.4,...,4.9}{
	\draw[black!40] (\x,0.7-0.2) -- (\x,5.1);
}
\draw[myred, dashed, line width=1.5] (1.5,1.5) -- (4.1,1.5) -- (4.1,4.1) -- (1.5,4.1) -- cycle;
\draw[myblue, line width=1.5] (1.4,3.5) -- (2.1,3.5) -- (2.1,4.2);
\draw[myblue, line width=1.5] (2.1,1.4) -- (2.1,2.1) -- (2.8,2.1) -- (2.8,2.8) -- (3.5,2.8) -- (3.5,1.4);
\draw[mygreen, line width=1.5] (1.4,3.5) -- (0.7,3.5) -- (0.7,4.9) -- (2.1,4.9) -- (2.1,4.2);
\draw[mygreen, line width=1.5] (2.1,1.4) -- (2.1,0.7) -- (3.5,0.7) -- (3.5,1.4);
\node[myblue] at (2.9,3.3) {$g_A$};
\node[mygreen] at (1.4,0.9) {$g_{\bar A}$};
\end{tikzpicture}
\end{equation*}
The reduced density matrix can be written as
\begin{align}
    \varrho_A(\mu) &= \Tr_{\bar{A}} \kket{\rho(\mu)}\bbra{\rho(\mu)}  \nonumber \\
    &= \frac{1}{\calZ(\mu)} \sum_{b, c} \sum_{\substack{g_A,\tilde{g}_A,\\\partial g_A = \partial \tilde{g}_A = b}} \sum_{\substack{
    h_A, \tilde{h}_A, \\\partial h_A = \partial \tilde{h}_A = c}} \Omega_{\bar{A}}(0; b)\Omega_{\bar{A}}(\mu; c)  e^{-2\mu|h_A| - 2\mu|\tilde{h}_A|} \kket{g_A, h_A}\bbra{\tilde{g}_A, \tilde{h}_A},
\end{align}
where $\Omega_{\bar{A}}(0;b)$ and $\Omega_{\bar{A}}(\mu;c)$ are the partition function of $g$ and $h$ loops in region $\bar{A}$ with a given boundary condition $b$ and $c$, respectively, 
\begin{align}
    \Omega_{\bar{A}}(\mu;c) = \sum_{h_{\bar{A}}, \, \partial h_{\bar{A}} = c} e^{-4\mu|h_{\bar{A}}|}.
\end{align}
The normalization factor $\calZ(\mu) = \Omega(0) \Omega(\mu)$ can be written accordingly as
\begin{align}
     \Omega(\mu) = \sum_c \Omega_{A}(\mu;c) \Omega_{\bar{A}}(\mu;c).
\end{align}

Next, we write down the $n$-th moment of the reduced density matrix
\begin{align}
\Tr \varrho_A^n(\mu) = \Gamma_{A}(0) \Gamma_{A}(\mu),
\end{align} 
where $\Gamma_{A}(0)$ and $\Gamma_A(\mu)$ are associated with $g$ and $h$ loops, respectively, and
\begin{align}
    \Gamma_{A}(\mu) = \frac{1}{\Omega^n(\mu)}\sum_{c}\Omega_{A}^n(\mu;c) \Omega_{\bar{A}}^n(\mu;c).\label{eq:n_moment}
\end{align}
Equation~\eqref{eq:n_moment} can be viewed as a ratio between two partition functions.
The denominator represents the partition function of $n$ copies of loop models, while the loop model in the numerator is further subject to the constraint that loops in different copies have the same boundary condition $c$.

The R\'enyi entropy therefore can also be interpreted as the excess free energy related to the above constraint
\begin{align}
S_A^{(n)} = \frac{1}{1-n} \log \Tr\varrho_A^n(\mu) = \frac{1}{n-1}\left( F_A^{(n)}(0) - nF_0(0)\right) + \frac{1}{n-1}\left( F_A^{(n)}(\mu) - nF_0(\mu)\right)
\end{align}
where $F_0(\mu) = -\log\Omega(\mu)$ is the free energy of the loop model with tension $\mu$, $F_A^{(n)}(\mu) = -\log \sum_{c}\Omega_A^n(\mu;c)\Omega_{\bar{A}}^n(\mu;c)$ is the free energy of $n$ copies of the loop model subject to the constraint.

The scaling of the excess free energy is easier to analyze when formulating the loop model in terms of Ising spins.
We again identify the loop configuration with Ising domain walls
\begin{align}
g^{(s)}_l = \frac{1 - \sigma_i^{(s)}\sigma_j^{(s)}}{2}, \quad h_l^{(s)} = \frac{1 - \sigma'^{(s)}_i\sigma'^{(s)}_j}{2},\quad s = 1,2,\cdots, n
\end{align}
Furthermore, we redefine the Ising variables
\begin{align}
    \sigma_i^{(s)} := \sigma_i^{(1)}\tau_i^{(s-1)}, \quad \sigma'^{(s)}_i := \sigma'^{(1)}_i\tau'^{(s-1)}_i,\quad s = 2, 3,\cdots, n.
\end{align}
The constraints on the loop configurations require $n$ copies of the Ising spins, $\sigma^{(s)}$ and $\sigma'^{(s)}$, on the boundary of the region $A$ having the same domain wall configurations.
Thus, they require the relative Ising spin $\tau^{(s)}_i$ and $\tau'^{(s)}_i$ having no domain wall on the boundary of $A$, i.e. pointing in the same direction.
There are $(n-1)$ such constraints in $F_A^{(n)}(\mu)$, therefore $F_A^{(n)}(\mu) = nF_0(\mu) + (n-1)\Delta F(\mu)$, where $\Delta F$ is the excess free energy for aligning one copy of the $\tau$ spins on the boundary of $A$. 
Correspondingly, the R\'enyi entropy $S_A^{(n)} = \Delta F(0) + \Delta F(\mu)$.

The excess free energy has two contributions.
The energetic part is always proportional to $|\partial A|$.
The entropic part counts the loss of degrees of freedom due to spin aligning.
In the ferromagnetic phase, the spins can fluctuate below the scale of correlation length $\xi$. 
Thus, $\Delta F \propto |\partial A|/\xi$ exhibits a strict area-law scaling.
In the paramagnetic phase, the spins can both fluctuate freely above the scale of correlation length $\xi$.
At the leading order, the free energy cost is proportional to $|\partial A|$.
However, the aligned spin can fluctuate globally giving rise to a subleading correction, i.e. $\Delta F = a|\partial A| - \log 2$.
Thus, the subleading term sharply changes at the ferromagnetic transition.

The above analysis indicates the R\'enyi entropy exhibits an area-law scaling $S_A^{(n)} = a |\partial A| - \gamma^{(n)}$, and the topological term $\gamma^{(n)}$ distinguishes two phases of the effective spin model.
When $\mu < \mu_c$, both spin models for $\sigma$ and $\sigma'$ are in the paramagnetic phase, giving rise to $\gamma^{(n)} = 2\log 2$.
When $\mu > \mu_c$, the spin model for $\sigma'$ (i.e. the loop model for $h$) is in the ferrmagnetic phase, reducing $\gamma^{(n)}$ to $\log 2$.
The result holds for an arbitrary R\'enyi order $n$ and also in the replica limit $n \to 1$.
Hence, the TEE of the EFD state sharply changes from $2\log 2$ to $\log 2$ when $m\bar{m}$ condenses at $p_c^{(2)}$.

We make an important remark that the effective model for the R\'enyi-$n$ entropy of the EFD is merely $n$ decoupled loop models.
The partition function $\calZ^n(\mu) = (\tr\rho^2)^n$.
Therefore, the transition occurs at the same critical threshold $p_c^{(2)}$ regardless of $n$.
This is fundamentally different from the transition in the $n$-th moment $\tr \rho^n$ of the corrupted density matrix, which occurs at a different critical threshold $p_c^{(n)}$.

The Toric code state subject to Pauli-Y and Z errors can be similarly analyzed.
These errors can induce a condensation of $f\bar{f}$ and $e\bar{e}$ in the EFD at the same critical threshold $p_c^{(2)}$.
We also note that, paired anyons $\alpha\bar{\alpha}$ and $\beta\bar{\beta}$ are mutual bosons and therefore can condense independently.
When two types of anyons are condensed, e.g. $e\bar{e}$ and $m\bar{m}$, the rest one $f\bar{f}$ is also condensed as $e \times m = f$.
The resulting EFD no longer contains condensed loop objects and will exhibit a vanishing TEE.

\section{Constraint on the phases in the EFD due to positivity}\label{sec:positivity}
In this section, we derive a constraint on the possible anyon condensates in the EFD $\kket{\rho}$ due to the positivity of the density matrix $\rho$.
We show that the composite object $\alpha\bar{\beta}$ can condense in $\kket{\rho}$ only if the anyon $\alpha\bar{\alpha}$, $\alpha\bar{\beta}$, and $\beta\bar{\beta}$ do not have mutual statistics.
This constraint does not affect the classification of the phases induced by incoherent errors studied in the main text.
Those phases feature condensed paired anyon $\alpha\bar{\alpha}$ in the EFD and do not violate this constraint.
The constraint plays a role when analyzing the possible anyon condensates induced by the coherent errors.
A similar constraint on the symmetry-protected topological phases in 1D mixed states due to positivity is derived in Ref.~\cite{Duivenvoorden:2017cha}.

We here provide a derivation.
Suppose $\alpha\bar{\beta}$ is condensed in $\kket{\rho}$, the open string operator $W_{\alpha\bar{\beta}}(\calP)$ along a path $\calP$ acquires a non-vanishing expectation value when the endpoints of $\calP$ are faraway, i.e.
\begin{align}
    \left|\langle\!\langle W_{\alpha\overline{\beta}}(\calP)\rangle\!\rangle\right| = \left|\frac{\bbra{\rho} W_{\alpha\overline{\beta}}(\calP) \kket{\rho}}{\bbrakket{\rho}{\rho}}\right| = \left|\frac{\tr\rho\, w_\alpha(\calP)\, \rho\, w_\beta^\dagger(\calP)}{\tr \rho^2}\right| > 0.
\end{align}
Due to the positivity of density matrix $\rho$, one can define the square root of $\rho$ such that $\rho = \sqrt{\rho}\sqrt{\rho}$. We then have
\begin{align}
    0<\left|\langle\!\langle W_{\alpha\overline{\beta}}(\calP)\rangle\!\rangle\right|^2 &= \left|\frac{\tr \sqrt{\rho}\,w_\alpha(\calP)\, \sqrt{\rho}\, \sqrt{\rho}\, w_\beta^\dagger(\calP)\,\sqrt{\rho}}{\tr \rho^2}\right|^2 \nonumber \\
    &\leq \frac{\tr \sqrt{\rho}\,w_\alpha(\calP)\, \sqrt{\rho}\, \sqrt{\rho}\, w_\alpha^\dagger(\calP)\,\sqrt{\rho}}{\tr \rho^2} \,\frac{\tr \sqrt{\rho}\,w_\beta(\calP)\, \sqrt{\rho}\, \sqrt{\rho}\, w_\beta^\dagger(\calP)\,\sqrt{\rho}}{\tr \rho^2} \nonumber \\
    &= \langle\!\langle W_{\alpha\overline{\alpha}}(\calP)\rangle\!\rangle \langle\!\langle W_{\beta\overline{\beta}}(\calP)\rangle\!\rangle
\end{align}
Here, we use the Cauchy-Schwarz inequality in the second line.
The inequality indicates that $W_{\alpha\overline{\alpha}}$ and $W_{\beta\overline{\beta}}$ are also non-decaying, i.e. the paired anyons $\alpha\bar{\alpha}$ and $\beta\bar{\beta}$ are condensed.
In the case that $\alpha\bar{\alpha}$, $\alpha\bar{\beta}$, and $\beta\bar{\beta}$ have non-trivial mutual statistics, the condensate of this triad cannot exist.

\section{Mapping decoherence-induced phases to boundary anyon condensates}
In this section, we map the decoherence-induced phases in the EFD to the boundary phases of a topologically ordered system.
In Sec.~\ref{sec:path_integral}, we start with the path integral formulation of the EFD state.
In Sec.~\ref{sec:mapping}, we map the decoherence-induced phases to the boundary phases of topologically ordered systems.
In particular, we establish the classification scheme for the possible phases in Abelian topological order subject to incoherent errors.
In Sec.~\ref{sec:examples}, we carry out the classification in three examples, the Toric code, double semion, and $\nu = 1/3$ Laughlin state with incoherent errors.
The results are summarized in Table~I of the main text.
Section~\ref{sec:nth} generalizes the field theory description for the second moment to the $n$-th moment $\tr\rho^n$.
We remark that, for Abelian topological order with incoherent errors, the number of possible phases (characterized by boundary anyon condensation) is independent of $n$.
Hence, each phase of the EFD has a correspondence in the replica limit $n\to 1$.

\subsection{Path integral formulation of the EFD}\label{sec:path_integral}
Here, we develop a path integral formulation of the EFD $\kket{\rho}$. 
The double ground state $\ket{\Psi_0}\otimes\ket{\Psi_0^*}$ in the EFD can be prepared using an infinite imaginary time evolution from an arbitrary state, which is equivalently a Euclidean path integral in spacetime region $\tau < 0$.
The wavefunction component at $\tau = 0$ can be formally written as
\begin{equation}
\begin{aligned}
    \calZ_{\tau<0} &[\varphi, \bar{\varphi}] := \big( \bra{\varphi} \otimes \bra{\bar{\varphi}} \big) \big(\ket{\Psi_0} \otimes\ket{\Psi_0^*} \big) = \int\displaylimits_{(\varphi',\bar\varphi')|_{\tau = 0^-} = (\varphi,\bar\varphi)} \calD (\varphi',\bar{\varphi}') e^{-\calS(\varphi') - \calS^*(\bar\varphi')}
\end{aligned},
\end{equation}
where $\varphi$ and $\bar\varphi$ are the dynamical variables in the ket and bra Hilbert space, the path integral is defined in the past spacetime region $\tau<0$, $\calS(\varphi')$ and $\calS^*(\bar\varphi')$ are a pair of conjugate Euclidean actions that yield the ground states $\ket{\Psi_0}$ and $\ket{\Psi_0^*}$, respectively.
The EFD $\kket{\rho}$ is then obtained by turning on the error channel which couples the field $\varphi$ and $\bar\varphi$ at $\tau=0$ time slice locally [illustrated in Fig.~1(b) of the main text]. 

The norm $\bbrakket{\rho}{\rho}$ is accordingly written as a Euclidean path integral in the entire spacetime, i.e. two path integrals in the future ($\tau>0$) and past ($\tau<0$) glued at $\tau = 0$:
\begin{equation}
    \bbrakket{\rho}{\rho} = \int \calD(\varphi_p, \bar{\varphi}_p, \varphi_f, \bar{\varphi}_f) \, \calZ_{\tau>0}[\varphi_f, \bar\varphi_f] \calZ_{\tau<0}[\varphi_p, \bar\varphi_p] e^{- \calS_{int}(\varphi_f,\bar\varphi_f,\varphi_p,\bar\varphi_p)},
\label{eq:path_int}
\end{equation}
where $\varphi_{f(p)}$ and $\bar{\varphi}_{f(p)}$ denote the future (past) field variables at the time slice $\tau=0^+$ ($\tau = 0^-$).
The future and past fields are coupled by the error channel, and the coupling is formally written as
$$
\begin{aligned}
    e^{-\calS_{int}(\varphi_f,\bar\varphi_f,\varphi_p,\bar\varphi_p)} = \bbra{\varphi_f, \bar\varphi_f }\calN^\dag \calN \kket{ \varphi_p, \bar\varphi_p }\,.
\end{aligned}
$$ 
In the absence of errors, $\calN = \bbI$ and $e^{-\calS_{int}} = \delta_{\varphi_p,\varphi_f} \delta_{\bar{\varphi}_p,\bar{\varphi}_f}$ becomes a constraint.
We further remark that individual Kraus operators and $\calN$ are generally non-Hermitian.
However, $\calS_{int}$ only contains terms determined by the Hermitian combination $\calN^\dag \calN$, which guarantees $\bbrakket{\rho}{\rho}$ being a real number.

\subsection{Mapping to (1+1)D boundary phases}\label{sec:mapping}
In this subsection, we take advantage of the path-integral formulation and work with the topological quantum field theory (TQFT) describing the long-distance physics of topological orders.
The invariance of TQFT under the $\pi/2$ spacetime rotation allows mapping decoherence-induced phases in the EFD to $(1+1)$D quantum phases on the edge of topologically ordered systems.
Here, we consider specifically the Abelian topological order described by the Chern-Simons (CS) theory.
In the case of incoherent errors, we carry out a classification of the possible decoherence-induced phases using the effective theory on the edge.
We demonstrate our general ideas with three concrete examples, the Toric code, the double semion state, and the Laughlin state. 
The results are summarized in Table I of the main text.
We also comment on the possible phases in the case of coherent errors.

We consider the EFD for the Abelian topologically ordered ground state subject to incoherent errors.
The ground state $\ket{\Psi_0}$ is described by a $U(1)^M$ Chern-Simons theory with the following Lagrangian
\begin{equation}
    \calL_{CS}[a] = \frac{1}{4\pi} \sum_{I,J = 1}^{M} K_{IJ}\, a^I \wedge \rd a^J + \cdots \,,
\end{equation}
where $K$ is a symmetric $M \times M$ integer matrix and $a^{I=1,2,\ldots,M}$ are the $U(1)$ gauge fields, the ellipsis denotes parts that involve matter fields~\cite{Read:1990zz,Wen:1992uk,Frohlich:1991wb,Wen:1995qn}. 
Matter fields are gapped and do not enter the dynamics at long distances, yet, they are necessary in giving rise to the anyonic excitations.
Accordingly, $\ket{\Psi_0^*}$ carries the conjugate topological order and is described by the same Lagrangian except that the $K$-matrix has an additional minus sign. To better distinguish them, we use $\bar{a}^{I=1,2,\ldots,M}$ to denote the gauge fields for $\ket{\Psi_0^*}$. 
We can combine these two parts and write the path integral $\calZ_{\tau<0}$ or $\calZ_{\tau>0}$ in terms of the following Lagrangian
\begin{equation}
    \calL_{CS}[\mathbf{a}] = \frac{1}{4\pi} \sum_{I,J = 1}^{2M} \mathbb{K}_{IJ}\, \mathbf{a}^I \wedge \rd \mathbf{a}^J + \cdots,
\end{equation}
where $\mathbb{K} = K \oplus (-K)$, and $\mathbf{a} = [a, \bar{a}]^T$ is a $2M$ component vector of the gauge fields in the Hilbert space $\calH\otimes \calH$.

Studying the decoherence-induced phases and the potential phase transitions in the EFD in the current form is not convenient.
On one hand, in general, anyonic excitations created by local errors explicitly involve matter fields and are complicated to analyze.
On the other hand, the transition, which happens at the temporal interface, is different from a ground state problem and largely unexplored.

These technical difficulties necessitate our following trick.
We perform a spacetime rotation to exchange the imaginary time $\tau$ and the spatial coordinate $x$ [Fig.~1(c) of the main text], i.e.,
\begin{equation}
    \tau \rightarrow  -x\,,\quad
    x \rightarrow \tau\,,\quad
    y\rightarrow y\,,
\end{equation}
where $x,y$ denote the spatial coordinates.
The rotation turns the temporal interface at $\tau = 0$ into a spatial one at $x = 0$.
The original path integral \eqref{eq:path_int} can be formally rewritten as
\begin{equation*}
\bbrakket{\rho}{\rho} = \int \calD(\varphi_L, \bar{\varphi}_L, \varphi_R, \bar{\varphi}_R) \, \calZ_{x>0}[\varphi_R, \bar\varphi_R] \calZ_{x<0}[\varphi_L, \bar\varphi_L] e^{- \tilde \calS_{int}(\varphi_R,\bar\varphi_R,\varphi_L,\bar\varphi_L)}\,,
\end{equation*}
where $\calZ_{x<0}$, obtained by rotating $\calZ_{\tau<0}$, is the path integral on the left side of the spatial interface $x = 0^-$, and $(\varphi_L,\bar{\varphi}_L)$ specify the value of the fields at the spatial interface. The meaning of $\calZ_{x>0}$ and $(\varphi_R,\bar{\varphi}_R)$ is similar.
$\tilde \calS_{int}$ describes the coupling between the two half systems at the spatial interface and generally takes a different form from the original one $\calS_{int}$.

After the rotation, the path integral is translationally invariant in the new time direction and describes the norm of a $(2+1)$D quantum ground state [Fig.~1(c) of the main text].
In this rotated picture, the decoherence channel introduces a 1D defect at $x = 0$ that couples the two sides, and the decoherence-induced phases correspond to the $(1+1)$D phases of the 1D defect.

However, it is not obvious that the evolution in the new time direction is generated by a Hermitian Hamiltonian.
In what follows, we will show that for the case of incoherent errors considered in this work, the path integral indeed describes the ground state of a Hermitian Hamiltonian.
The incoherent channel takes the following form when acting on the EFD,
\begin{align}
    \calN = \prod_i \calN_i, \quad \calN_i = \sum_k \calK_{k,i}\bar{\calK}_{k,i} = \sum_k |c_{k}|^2\calA_i[\alpha_k]\bar{\calA}_i[\alpha_k],
\end{align}
where the Kraus operator $\calK_{k,i} = c_k \calA_i[\alpha_k]$ for incoherent error creates a cluster of anyons labeled by $\alpha_k$ in the vicinity of site $i$.
For example, it can create anyon $\alpha_k$ and its anti-particle on neighboring sites.
A key observation is that $\calK_{k,i}\bar{\calK_{k,i}}$ only creates or hops paired anyons $\alpha_k\bar{\alpha}_k$ in the EFD with a positive real weight $|c_k|^2$.

The Lagrangian that describes the 1D defect has three contributions
\begin{align}
    \calL = \calL_0 + \calL_1 + \calL_{\calN},
\end{align}
where $\calL_0$ describes the low energy edge excitations for the $(2+1)D$ theory in the region $x > 0$ and $x < 0$.
The interaction on the defect is derived from $\tilde{S}_{int}$ and contains two parts.
The first part originates from the definition of path integral and couples degrees of freedom on the left and the right given by $\calL_1$, which is present even in the absence of a quantum channel.
The second part $\calL_\calN$ is introduced by the quantum channel.
In what follows, we explicitly write down the three terms in the Lagrangian and analyze the possible phases.

We start with the low energy edge theory $\calL_0$ for the CS theory in the bulk.
We consider how the path integral for the past $\calZ_{\tau<0}$ changes under the rotation:
\begin{itemize}
\item Originally, $\calZ_{\tau<0}[\varphi_p, \bar{\varphi}_p]$ describes the path integral that prepares the ground state of a system on a whole plane, where $\varphi_p, \bar{\varphi}_p$ specifies the component of the wavefunction. Note that $\varphi_p$ and $\bar{\varphi}_p$ includes both the CS gauge fields and the matter fields, and thus they are complicated to deal with.

\item The spacetime rotation transforms it into $\calZ_{x>0}[\varphi_R, \bar{\varphi}_R]$, a path integral defined on the half-space but the entire time domain, and $\varphi_R, \bar{\varphi}_R$ specifies the boundary value of the fields.
The bulk part of the path integral contains the same CS action as that before the rotation and the part involving matter fields.
The boundary has a simple description in terms of compact bosons, which is the main reason why the rotation trick can be useful.

\item From now on, we will replace $\varphi_R$ and $\bar{\varphi}_R$ by the compact boson fields $\phi_R, \bar{\phi}_R$, which correspond to the bulk gauge fields $a, \bar{a}$. 
Then, the low-energy dynamics of the boundary of the CS theory is governed by the following effective Lagrangian
\begin{equation}
    \calL_R[\Phi_R] = \frac{1}{4\pi} \sum_{I,J=1}^{2M} -\mathbb{K}_{IJ} \ri \partial_\tau \Phi_R^I \partial_y \Phi_R^J - \mathbb{V}_{IJ} \partial_y \Phi_R^I \partial_y \Phi_R^J,
\end{equation}
where $\Phi_R := [\phi_R, \bar{\phi}_R]$, $\mathbb{K}$ is the same K-matrix as the one for the bulk CS action, and $\mathbb{V}$ is a positive semi-definition matrix that determines the velocity of the boson fields.
The path integral with fixed boundary values of the boson fields is then determined by $\calL_R[\Phi_R]$ classically:
\begin{equation}
    \calZ_{x<0}[\Phi_R] = \exp\left( - \int \rd\tau \rd y \calL_R[\Phi_R] \right)\,.
\end{equation}
\end{itemize}
The transformation of the path integral for the future $\calZ_{\tau>0}$ can be analyzed similarly. The result is related to what we have obtained above by a spatial reflection
\begin{equation*}
    \calL_L[\Phi_L] = \frac{1}{4\pi} \sum_{I,J=1}^{2M} \mathbb{K}_{IJ} \ri \partial_\tau \Phi_L^I \partial_y \Phi_L^J - \mathbb{V}_{IJ} \partial_y \Phi_L^I \partial_y \Phi_L^J,
\end{equation*}
where $\Phi_L := [\phi_L, \bar{\phi}_L]$, and the minus sign in front of the $K$-matrix comes from the spatial reflection.
The path integral $\calZ_{x>0}[\Phi_L]$ can also be determined classically
\begin{equation}
    \calZ_{x>0}[\Phi_L] = \exp\left( - \int \rd\tau \rd y \calL_L[\Phi_L] \right)\,.
\end{equation}

We combine the low-energy excitations from the left and the right and obtain the Lagrangian describing the edge of the quadruple topological order
\begin{equation}
    \mathcal{L}_0[\phi] = \frac{1}{4\pi} \sum_{I,J} \mathbb{K}_{IJ}^{(2)} \ri \partial_\tau \phi^I \partial_y \phi^J - \mathbb{V}_{IJ}^{(2)} \partial_y \phi^I \partial_y \phi^J\,,\label{eq:edge_lagrangian}
\end{equation}
where the $4M\times 4M$ $K$-matrix $\mathbb{K}^{(2)} = \mathbb{K} \oplus \mathbb{K}$, velocity matrix $\mathbb{V}^{(2)} = \bbV\oplus\bbV$, and $\phi := [\bar{\phi}_R, \phi_R, \phi_L, \bar{\phi}_L]$.

Next, we discuss the interaction in the effective theory.
The first part $\calL_1$ is associated with the definition of path integral and couples $\phi_L$ to $\phi_R$ and $\bar{\phi}_L$ to $\bar{\phi}_R$.
It leads to a gapped phase of the defect in the absence of errors.
The second part $\calL_\calN$ is introduced by the error channel.
It couples the bosonic fields from the same copy of the density matrix (or equivalently EFD), namely $\phi_s$ to $\bar{\phi}_s$ for $s = L, R$.

To explicitly write down the interaction in terms of the compact bosons, we first list the excitations in the effective theory $\calL_0$.
The excitations are labeled by integer vectors $\bm{l}_\alpha$ and are created by $e^{\ri\bm{l}_\alpha^T\cdot \phi}$.
They have anyonic statistics, and the braiding of two anyons $\bm{l}_\alpha$ and $\bm{l}_\beta$ yields a phase 
\begin{equation}
\theta_{\alpha\beta} = 2\pi \bm{l}_\alpha^T (\mathbb{K}^{(2)})^{-1} \bm{l}_\beta.
\end{equation}
We remark that, in general, a single anyon cannot be created using a local operator.
Instead, a pair of anyon $\alpha$ and its antiparticle $\alpha'$ can be created using a local operator, which is described by $\Psi_{\Lambda} := e^{\ri (\mathbb{K}^{(2)} \Lambda)^T \cdot \phi}$ with $\Lambda$ being an integer vector.
Such excitation has trivial braiding with other excitations, i.e. $\theta_{\alpha, \mathbb{K}^{(2)}\Lambda} = 0 \mod 2\pi, \forall \alpha$.

Furthermore, the form of the interaction is constrained by the global $\calG^{(2)} = \mathbb{Z}_2 \times \mathbb{Z}_2^\mathbb{H}$ symmetry of the 1D defect.
The first $\mathbb{Z}_2$ is an anti-unitary symmetry generated by the hermitian conjugation in the double Hilbert space.
The path integral represents the norm of the EFD, which is a real number.
Thus, the effective theory is invariant under the transformation
\begin{equation}
\begin{aligned}
    \phi_{L}^I &\leftrightarrow \phi_{R}^I, \quad \bar{\phi}_{L}^I \leftrightarrow \bar{\phi}_{R}^I.
\end{aligned}
\end{equation}
In addition, the EFD represents the density matrix, which is hermitian in the original Hilbert space.
Thus, edge theory exhibits another anti-unitary $\mathbb{Z}_2^\mathbb{H}$ symmetry due to the hermiticity of the density matrix, which acts as
\begin{equation}
\begin{aligned}
    \phi_{s}^I &\leftrightarrow \bar{\phi}_{s}^I,\quad s = L,R.
\end{aligned}
\end{equation}
We remark that the K-matrix $\mathbb{K}^{(2)}$ flips the sign under the anti-unitary symmetry, however, the Lagrangian $\calL_0$ is invariant as the minus sign cancels that from the imaginary identity.

We are now ready to explicitly write down the interaction in terms of edge excitations. 
The first part $\calL_1$ from the definition of path integral can be written in terms of the back-scattering between the left and the right fields.
It is generally given by a hopping term between the left and the right
\begin{align}
    \sum_{\Lambda} \frac{B_{\Lambda}(y)e^{\ri \theta(y)}}{2} \left(\Psi_{L,\Lambda}^\dagger \Psi_{R,\Lambda} + \bar{\Psi}_{R,\Lambda}^\dagger \bar{\Psi}_{L,\Lambda}\right) + h.c.,
\end{align}
where $\Psi_{L,\Lambda} := e^{\ri (K\Lambda)^T \phi_L}$, $\Psi_{R,\Lambda} := e^{-\ri (K\Lambda)^T \phi_R}$, $\bar{\Psi}_{L,\Lambda} := e^{\ri (-K\Lambda)^T \bar{\phi}_L}$, and $\bar{\Psi}_{R,\Lambda} := e^{\ri (K\Lambda)^T \bar{\phi}_R}$ are local operators on the edge, $\Lambda$ is a $M$-component integer vector, the real coefficient $B_\Lambda$ and phase factor $\theta$ have spatial dependence in general.
In terms of the compact boson fields, the back-scattering takes the form
\begin{equation}
    \mathcal{L}_1 = \sum_\Lambda B_\Lambda \Big[ \cos\Big(\sum_{I = 1}^M (K\Lambda)_I (\phi_L^I + \phi_R^I) +\theta \Big) + \cos\Big(\sum_{I = 1}^M (K\Lambda)_{I}(\bar{\phi}_L^I + \bar{\phi}_R^I)+\theta\Big)\Big].\label{eq:edge_int_ref}
\end{equation}
We remark that the coefficients for two cosine terms are forced to the same real number due to the global symmetry $\mathcal{G}^{(2)}$.
With large coefficients, such cosine terms tend to pin the bosonic fields $\phi+\bar{\phi}$ and lead to a gapped phase of the defect without any error channel.
These interaction terms only create anyon bound states on the left and the right in the form $\alpha_L\alpha'_R = e^{\ri\bm{l}^T\cdot(\phi_L + \phi_R)}$ and $\bar{\alpha}_L\bar{\alpha}'_R = e^{-\ri\bm{l}^T\cdot(\bar{\phi}_L + \bar{\phi}_R)}$.

The second part $\calL_\calN$ of the interaction is introduced by the quantum channel and couples the fields from the same copy of the density matrix.
For incoherent error channels, before the rotation, $\calN^\dagger \calN$ only creates paired anyons $\alpha\bar{\alpha}$ in the EFD.
Since $\calN^\dagger \calN$ is a Hermitian operator in the double space, an anyon cluster of $\alpha\bar{\alpha}$ and its Hermitian conjugate that involves its anti-particle $\alpha'\bar{\alpha}'$ are created with the same (positive real) weight.
Therefore, in the rotated picture, the quantum channel introduces local couplings $e^{\ri(K\Lambda)^T(\phi_s - \bar{\phi}_s)}$ and $e^{-\ri(K\Lambda)^T(\phi_s - \bar{\phi}_s)}$ ($s = L, R$) with the same real weights, which combine to the cosine terms at the 1D defect,
\begin{align}
    \mathcal{L}_{\mathcal{N}} = \sum_\Lambda C_\Lambda \sum_{s = L, R} \cos\left( \sum_{I = 1}^M (K\Lambda)_I (\phi_{s}^I - \bar{\phi}_{s}^I) \right),\label{eq:int_quantum_channel}
\end{align}
where $C_\Lambda$ is a real coefficient with spatial dependence.
The coefficients of the cosine terms for $s = L, R$ are the same due to the $\mathbb{Z}_2$ symmetry.

We note that the cosine terms in the Lagrangian all have real coefficients.
After the canonical quantization in the rotated picture, the edge Hamiltonian is Hermitian.
This guarantees that the decoherence-induced phases indeed map to ground-state phases on the edge.

The possible phases of the defect are classified by the inequivalent ways of gapping out $\calL_0$ in Eq.~\eqref{eq:edge_lagrangian} by condensing bosonic excitations.
For Abelian topological order, the bosonic objects form a group, dubbed the Lagrangian subgroup $\calM$~\cite{Levin:2013gaa,Barkeshli:2013,Wang:2012am}.
Thus, the phases of Abelian topological order subject to incoherent errors are classfied by the subgroup $\calM$ that satisfying the following criteria:
\begin{enumerate}
	\item $e^{\ri \theta_{\bm{m}\bm{m'}}} = 1, \forall \bm{m}, \bm{m'} \in \calM$;
	\item $\forall \bm{l} \notin \calM$, $\exists \bm{m}$ s.t. $e^{\ri \theta_{\bm{ml}}} \neq 1$;
        \item $\forall \bm{m} \in \calM$, $ g\bm{m} g^{-1} \in \calM, \forall g \in \calG^{(2)}$;
	\item (Incoherent error) $[\bm{1}, \bm{1},-\bm{1},-\bm{1}]^T \cdot \bm{m} = 0 \mod \mathbb{K}^{(2)}\Lambda, \forall \bm{m} \in \calM$,
\end{enumerate}
where $\Lambda$ is a $4M$-component integer vector, $\bm{1}$ is an $M$-component vector with each element being unity.
Here, the third criterion originates from the symmetry constraints, i.e. the excitations related by symmetry transformations must condense simultaneously.
The last criterion is due to incoherent errors.
The interaction on the edge only creates anyon bound states $\alpha_L\bar{\alpha}_L$, $\alpha_R\bar{\alpha}_R$, $\alpha_L\alpha'_R$, $\bar{\alpha}_L\bar{\alpha}'_R$ and their fusion results.
Therefore, the condensed objects satisfy the last criterion.

In the case of coherent errors, the phases are still classified by the Lagrangian subgroup.
However, since the Kraus operators can, for example, create anyon excitations $\alpha_s$ and $\bar{\alpha}_s$ in a single copy, the condensed objects need not satisfy the last criterion.
Correspondingly, in the Lagrangian, the coherent errors can introduce the cosine terms, $\cos((K\Lambda)^T \cdot \phi_{s})$ and $\cos((K\Lambda)^T \cdot \bar{\phi}_{s})$, which solely depends on the compact boson in an individual copy of the topological order.
This can possibly lead to additional decoherence-induced phases.

Each edge condensate corresponds to a distinct phase of the EFD, and its condensed objects also determine the capability of encoding information in the error-corrupted state.
In the main text, we mention that the logical operator $W_{\alpha\bar{\alpha}}(\ell)$ can no longer encode quantum information in the condensate of paired anyons $\alpha\bar{\alpha}$ in the EFD, which is equivalent to condensing $\alpha_s\bar{\alpha}_s$ with $s = L,R$ in the $(1+1)$D theory of the defect.
Hence, examining whether $\alpha_s\bar{\alpha}_s$ is condensed on the edge can determine the encoding properties in the error-corrupted state.

\subsection{Examples}\label{sec:examples}

Here, we classify the possible phases by enumerating the Lagrangian subgroups in three examples, 2D Toric code, double semion model, and $\nu = 1/3$ Laughlin state, subject to incoherent errors.
We comment on the information encoding in each phase.
The results are summarized in Table~I of the main text.

\emph{Example 1: Toric code.}---
The first example we consider is the 2D Toric code subject to incoherent errors.
We find five distinct gapped phases (see Table~I~\footnote{Here and in the following, Table~I refers to the table in the main text.}).

To start with, the low energy excitations in 2D Toric code are described by the CS theory with K-matrix
\begin{align}
    K_{TC} = \begin{pmatrix}
    0 & 2 \\
    2 & 0
    \end{pmatrix}.
\end{align}
The Toric code has four superselection sectors: $1, e, m, f$. Here, $e$ and $m$ are self-boson and exhibit mutual semionic statistics. The fermion $f$ is a composite object of $e$ and $m$.
These anyonic quasiparticles are labeled by integer vectors
\begin{align}
    \bm{l}_e = \begin{pmatrix}
    1 \\
    0
    \end{pmatrix}, \quad \bm{l}_m = \begin{pmatrix}
    0 \\
    1
    \end{pmatrix}, \quad \bm{l}_f = \begin{pmatrix}
    1 \\
    1
    \end{pmatrix}.
\end{align}

The first gapped phase is achieved when no error is present. 
In this case, the definition of path integral introduces the cosine terms in Eq.~\eqref{eq:edge_int_ref}
\begin{equation}
\begin{aligned}
    \calL_{1} = \sum_{I=1,2} B_I \Big[&\cos\left(2(\phi_L^I +  \phi_R^I ) + \theta\right) + (\phi_s \leftrightarrow \bar{\phi}_s) \Big] + \cdots.
\end{aligned}\label{eq:toric_code_int_ref}
\end{equation}
Here, the ellipses represent the less relevant terms.
With sufficiently large coefficients $B_I$, these cosine terms can induce an edge condensate of $e_{L}e_R, \bar{e}_L\bar{e}_R, m_Lm_R$, and $\bar{m}_L\bar{m}_R$ (Phase I in Table~I).
The corrupted density matrix in this phase can encode quantum information because $\alpha_s\bar{\alpha}_s$ is not condensed.

The incoherent error channel can induce distinct edge phases.
Here, we have three cosine terms in the form of Eq.~\eqref{eq:int_quantum_channel}
\begin{equation}
\begin{aligned}
    \mathcal{L}_{\mathcal{N},e} &= C_e\sum_{s = L, R} \cos\left(2(\bm{l}_e^T \cdot \phi_{s}-\bm{l}_e^T \cdot \bar{\phi}_{s})\right),\\
    \mathcal{L}_{\mathcal{N},m} &= C_m\sum_{s = L, R} \cos\left(2(\bm{l}_m^T \cdot \phi_{s}-\bm{l}_m^T \cdot \bar{\phi}_{s})\right),\\
    \mathcal{L}_{\mathcal{N},f} &= C_f\sum_{s = L, R} \cos\left(2(\bm{l}_f^T\cdot \phi_{s}-\bm{l}_f^T\cdot \bar{\phi}_{s})\right),
\end{aligned}\label{eq:toric_code_int_error}
\end{equation}
which are generated by the incoherent channels that create $e$, $m$, and $f$, respectively.
With a sufficiently large coefficient, any one of three cosine terms can develop a classical value leading to a condensate of $\alpha_s \bar{\alpha}_s$ with $\alpha = e, m, f$, which corresponds to Phase II-IV in Table~I, respectively.
In these phases, the remaining logical operators mutually commute, and the corrupted state is a classical memory.

Finally, when two cosine terms develop a classical value, the remaining one also achieves a classical value (Phase V in Table~I).
Here, the corrupted state is a trivial state, and one cannot encode any information.

Allowing coherent errors may lead to additional gapped phases beyond the above five possibilities.
For example, we consider the amplitude-damping channel
\begin{align}
\calK_{1,i} = \frac{1+\sqrt{1-\kappa}}{2}\bbI + \frac{1-\sqrt{1-\kappa}}{2} Z_i, \quad \calK_{2,i} =  \frac{\sqrt{\kappa}}{2}(X_i+\ri Y_i).
\end{align}
The channel can generate $e, \bar{e}, m, \bar{m}, e\bar{m}, m\bar{e}$ in the EFD, which cannot be created by incoherent errors.
Among these anyons, $e, \bar{e}, m, \bar{m}$ can condense according to the constraint of the positivity in Sec.~\ref{sec:positivity}.
In the effective theory, condensing a subset of them can lead to additional phases of the 1D defect labeled by the Lagrangian subgroups with the following generators:
\begin{itemize}
    \item $e_s, \bar{e}_s \text{ for } s = L, R$;
    \item $m_s, \bar{m}_s \text{ for } s = L, R$;
\end{itemize}
However, as the amplitude damping channel creates all these anyons together with the paired anyon $\alpha_s\bar{\alpha}_s$, these possible phases will compete.
It requires numerical simulation to determine which anyon condenses at low energy.
Also, it remains open whether a specific coherent error channel can realize any of these phases. 

We make a few remarks. 
First, the phase with $e$ or $m$ anyon condensed in individual copies can be obtained by an imaginary time evolution of the Toric code ground state.
For example, $\ket{\Psi_0} \rightarrow e^{\mu \sum_i Z_i}\ket{\Psi_0}$ can realize the edge condensate of $e_s$ and $\bar{e}_s$.
Second, the above two phases are studied in the context of 1D defect phases in double Toric code and are referred to as the ``cut-open" boundary~\cite{Lichtman:2020nuw}.

\emph{Example 2: Double semion.}---
The second example we consider is the double semion model subject to incoherent errors~\cite{LevinGu:2012}.
Again, we enumerate the possible Lagrangian subgroups that characterize the distinct phases.
We find five phases summarized in Table~I.

The low-energy theory for double semion topological order is the Abelian CS theory with K-matrix
\begin{align}
    K_{DS} = \begin{pmatrix}
    2 & 0 \\
    0 & -2
    \end{pmatrix}.
\end{align}
The double semion model has four superselection sectors: $1, m_a, m_b, b$, where $m_a$ is a semion, $m_b$ is an anti-semion, and the boson $b$ is a composite object of $m_a$ and $m_b$.
The semion $m_a$ and anti-semion $m_b$ are mutual boson.
These anyons are characterized by the integer vectors
\begin{align}
    \bm{l}_{m_a} = \begin{pmatrix}
    1 \\
    0
    \end{pmatrix}, \quad \bm{l}_{m_b} = \begin{pmatrix}
    0 \\
    1
    \end{pmatrix}, \quad \bm{l}_b = \begin{pmatrix}
    1 \\
    1
    \end{pmatrix}.
\end{align}

The double semion model on the torus has a four-fold degenerate ground state that can encode two logical qubits.
The logical operators are the loop operators $w_\alpha(\ell)$ transporting $\alpha = m_a, m_b$ along two inequivalent large loops $\ell = \ell_1,\ell_2$.
Distinct from the case of Toric code, the operators along different loops, $w_\alpha(\ell_1)$ and $w_\alpha(\ell_2)$ for $\alpha = m_a, m_b$, anti-commute due to the anyon self-statistics and therefore serve as Pauli-X and Z operators for a logical qubit.

The first gapped phase (Phase I in Table~I) of the defect is realized when no error is present.
The backscattering term $\calL_1$ condenses the anyon bound states $\alpha_L{\alpha}'_R$ and $\bar{\alpha}_L\bar{\alpha}'_R$ for $\alpha = m_a, m_b, b$.
This phase can encode two qubits of quantum information.

The incoherent channel in the double semion model can introduce the following cosine terms:
\begin{equation}
\label{eq:double_semion_int_error}
\begin{aligned}    
\mathcal{L}_{\mathcal{N},m_a} &= C_{m_a}\sum_{s = L, R} \cos\left(2(\bm{l}_{m_a}^T\cdot\phi_{s}-\bm{l}_{m_a}^T\cdot\bar{\phi}_{s})\right), \\
\mathcal{L}_{\mathcal{N},m_b} &= C_{m_b}\sum_{s = L, R} \cos\left(2(\bm{l}_{m_b}^T\cdot\phi_{s}-\bm{l}_{m_a}^T\cdot\bar{\phi}_{s})\right),\\
\mathcal{L}_{\mathcal{N},b} &= C_b\sum_{s = L, R} \cos\left(2(\bm{l}_{b}^T\cdot\phi_{s}-\bm{l}_{b}^T\cdot\bar{\phi}_{s})\right),
\end{aligned}
\end{equation}
which creates paired anyons $\alpha_s\bar{\alpha}_s$ for $\alpha = m_a, m_b, b$, respectively.
Condensing each one of these terms induces Phase II-IV in Table~I, respectively.
In Phase II [III], one can encode information in $w_{m_b}(\ell)$ [$w_{m_a}(\ell)$] for $\ell = \ell_1,\ell_2$.
These logical operators do not mutually commute and therefore can encode one qubit of quantum information.
In Phase IV, the condensation of $b_s\bar{b}_s$ also results in a classical memory~\footnote{We appreciate Zhuan Li and Roger Mong for pointing out a mistake in the early version of our manuscript regarding information encoding in the corrupted double semion states.}. 
To understand this, we first recall the logical operators in the double-semion state on the torus, which encodes two-qubit quantum information. 
The operators $w_{m_a}(\ell)$ along non-contractible loops $\ell_1$ and $\ell_2$ anti-commute and serve as the logical Pauli-X and Z operator for the first qubit, i.e. $X_1 := w_{m_a}(\ell_1)$ and $Z_1 := w_{m_a}(\ell_2)$.
The operator $w_{m_b}(\ell_1)$ and $w_{m_b}(\ell_2)$ are identified with the logical operators for the second qubit, i.e. $X_2 := w_{m_b}(\ell_1)$ and $Z_2 := w_{m_b}(\ell_2)$.
Condensing $b\bar{b}$ in the doubled state leads to the logical errors $w_b(\ell_1) = X_1X_2$ and $w_b(\ell_2) = Z_1Z_2$ and completely dephase the logical qubits.
The resulting logical density matrix is diagonal in the Bell-state basis, therefore only encodes classical information.
Condensing two terms in Eq.~\eqref{eq:double_semion_int_error} indicates the third is condensed and results in Phase V, which cannot encode information.

Introducing coherent errors can potentially realize an additional phase by condensing the boson in individual copies, $b_s$ and $\bar{b}_s$.
Again, designing a specific coherent error channel to possibly realize this phase is left for future study.

\emph{Example 3: $\nu = 1/3$ Laughlin state.---}
The third example is $\nu = 1/3$ Laughlin state~\cite{Laughlin1983} with incoherent errors.
In this case, we find two phases summarized in Table~I.

The low-energy excitations in the Laughlin state is described by the Abelian CS theory with K-matrix $K_{1/3} = (3)$.
It contains three superselection sectors $\{1, \eta, \eta^2\}$.
$\eta$ and $\eta^2$ are the quasiparticle and quasihole, and they are labeled by the charge vector $\bm{l}_\eta = (1)$ and $\bm{l}_{\eta^2} = (2)$ and carry charge $e/3$ and $-e/3$, respectively.

There is only one type of incoherent channel; its Kraus operators create a pair of quasiparticle and quasihole.
When the error is below the threshold, the corrupted state is a quantum memory characterized by the Lagrangian subgroup generated by $\eta_L \eta_R^2$ and $\bar{\eta}_L\bar{\eta}_R^2$ (Phase I in Table~I).
When error proliferates, we obtain a trivial phase with a subgroup generated by $\eta_L \bar{\eta}_L$ and $\eta_R \bar{\eta}_R$ (Phase II in Table~I).

\subsection{Generalization to the $n$-th moment}\label{sec:nth}
\begin{figure}
\centering
\begin{tikzpicture}
\pgftext{\includegraphics[width=0.7\textwidth]{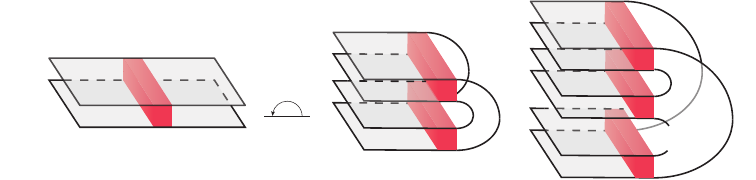}} at (0pt,0pt);
\node at (-5.5, 0.96) {(a)};
\draw [->,>=stealth] (-3.6, 0.7) node[above]{1D defect} -- (-3.7,0.4);
\node at (-3.6, -0.9) {2 $\times$ (topo. order)};
\node at (-1.0, 0.96) {(b)};
\node at (-1.5, 0.05) {\scriptsize Folding};
\node at (0.3, 0.4) {\scriptsize $\phi_1$};
\node at (0.3, 0) {\scriptsize $\bar{\phi}_1$};
\node at (0.3, -0.4) {\scriptsize $\phi_2$};
\node at (0.3, -0.8) {\scriptsize $\bar{\phi}_2$};
\node at (2.3, 1.4) {(c)};
\node at (3.5, 0.85) {\scriptsize $\phi_1$};
\node at (3.5, 0.5) {\scriptsize $\bar{\phi}_1$};
\node at (3.5, 0.05) {\scriptsize $\phi_2$};
\node at (3.5, -0.3) {\scriptsize $\bar{\phi}_2$};
\node at (4.9, -0.7) {\scriptsize $\vdots$};
\node at (3.5, -0.93) {\scriptsize $\phi_n$};
\node at (3.5, -1.28) {\scriptsize $\bar{\phi}_n$};
\end{tikzpicture}
\caption{(a) The path integral represents the norm of the ground state of double 2D topological order with 1D defect (red region). (b) Folding the double topological order around the 1D defect results in quadruple topological order with a 1D boundary. The red regions represent the coupling $\calL_\calN$ induced by the quantum channel. The couplings between $\phi_1$ and $\bar{\phi}_2$ (also $\bar{\phi}_1$ and $\phi_2$) are denoted by $\calL_1$ inherited from the definition of path integral. (c) The edge theory associated with the $n$-th moment $\tr\rho^n$ of the density matrix. The generalized swap interaction $\calL_1^{(n)}$ couples the compact bosons in the neighboring copies, i.e. $\bar{\phi}_{s}$ and $\phi_{s+1}$ for $s = 1, 2,\cdots, n$.}
\label{suppfig:nth_moment}
\end{figure}

So far, the analysis focuses on the distinct phases in the EFD, which pertains to analyzing the second moment of the density matrix, i.e. the norm of the EFD $\bbrakket{\rho}{\rho} = \tr \rho^2$.
In this subsection, we generalize the problem to the $n$-th moment $\tr\rho^n$ and map the decoherence-induced phases to $(1+1)$D boundary phases.

Before discussing the $n$-th moment, we develop an equivalent formulation of the second moment $\tr \rho^2$.
The phases of the $(1+1)$D defect can be converted to the $(1+1)$D edge phases by folding around the defect as shown in Fig.~\ref{suppfig:nth_moment}(b)~\cite{BeigiShor2011,Kapustin2011}.
In the folded picture, we have four copies of the topological order.
On the edge, we relabel the bosonic fields as
\begin{align}
    \phi := [\bar{\phi}_R, \phi_R, \phi_L, \bar{\phi}_L] = [\phi_1, \bar{\phi}_1, \phi_2, \bar{\phi}_2],
\end{align}
where $\phi_{1(2)}$ and $\bar{\phi}_{1(2)}$ denote the fields associated with the copy-$1(2)$ of the density matrix.

The path integral describing the ground state of quadruple topological order with an edge corresponds to an overlap in the quadruple Hilbert space $\calH^{\otimes 4}$ before the rotation
\begin{align}
\tr\rho^2 = \bbra{\mathbb{C}^{(2)}}\left(\kket{\rho}\otimes \kket{\rho}\right) := \int \mathcal{D}(\varphi_1, \varphi_2) \left(\bra{\varphi_1}\otimes \bra{\varphi_2}\right)\otimes \kket{\rho} \, \left(\bra{\varphi_2}\otimes \bra{\varphi_1}\right)\otimes \kket{\rho},
\end{align}
where $\bbra{\mathbb{C}^{(2)}}$ is a ``swap" reference state that couples the fields in the ket and bra Hilbert space of the different copies of the EFD.

The overlap is readily generalized to describe the $n$-th moment $\tr\rho^n$.
Here, in $2n$ copies of the Hilbert space $\calH^{\otimes 2n}$, we consider 
\begin{align}
\tr\rho^n = \bbra{\mathbb{C}^{(n)}}\left(\kket{\rho}^{\otimes n}\right) := \int \mathcal{D}(\varphi_1, \varphi_2, \cdots, \varphi_n) \prod_{s = 1}^n \left(\bra{\varphi_s}\otimes \bra{\varphi_{s+1}}\right)\otimes \kket{\rho},
\end{align}
where $\varphi_{n+1} := \varphi_1$, and $\bbra{\mathbb{C}^{(n)}}$ is a generalized swap reference state.

We can similarly develop a path integral formulation of the overlap.
In the rotated picture, it corresponds to the ground state of $2n$ copies of the topological order with an $(1+1)$D edge.
The effective edge theory also contains three parts
\begin{equation}
\calL^{(n)} = \calL_0^{(n)} + \calL_1^{(n)} + \calL_\calN^{(n)}.
\end{equation}
The first part $\calL_0^{(n)}$ describes the low-energy excitations in $2n$ copies of the topological order
\begin{align}
\calL_0^{(n)}[\phi] = \frac{1}{4\pi} \sum_{I,J} \mathbb{K}_{IJ}^{(n)} \ri\partial_\tau \phi^I \partial_y \phi^J - \mathbb{V}_{IJ}^{(n)} \partial_y \phi^I \partial_y \phi^J
\end{align}
where the compact bosonic field $\phi := [\phi_1, \bar{\phi}_1, \phi_2, \bar{\phi}_2, \cdots, \phi_n, \bar{\phi}_n]$ is a $2nM$-component vector, $\mathbb{K}^{(n)} = [K \oplus (-K)]^{\oplus n}$, and $\mathbb{V}^{(n)} = V^{\oplus 2n}$.
The second part $\calL_1^{(n)}$ is the interaction introduced by the reference state, which takes a general form
\begin{align}
\calL_1^{(n)} = \sum_\Lambda B_\Lambda \sum_{s = 1}^n \cos\left( \sum_I (K\Lambda)_I (\phi_s^I + \bar{\phi}_{s+1}^I) + \theta \right).
\end{align}
where $\bar{\phi}_{n+1} := \bar{\phi}_1$.
The last part $\calL_\calN^{(n)}$ describes the error channel. The incoherent errors can introduce the following terms similar to Eq.~\eqref{eq:int_quantum_channel}
\begin{align}
\calL_\calN^{(n)} = \sum_\Lambda C_\Lambda \sum_{s = 1}^n \cos\left( \sum_I (K\Lambda)_I (\phi_s^I - \bar{\phi}_s^I)\right).
\end{align}

The edge theory exhibits a global symmetry $\mathcal{G}^{(n)} = \mathbb{Z}_n \times \mathbb{Z}_2^\mathbb{H}$.
We note that the $\mathbb{Z}_n$ symmetry is a unitary symmetry generated by the cyclic permutation over $n$ copies of the density matrix
\begin{align}
    \phi_s \leftrightarrow \phi_{s+1}, \quad \bar{\phi}_s \leftrightarrow \bar{\phi}_{s+1}.
\end{align}
In the case of $n = 2$, $\calG^{(n)}$ reduces to $\calG^{(2)}$ in Sec.~\ref{sec:mapping} as the permutation over two copies of the density matrix is a combination of the Hermitian conjugation of the EFD and that of the density matrix.

We remark that the incoherent errors can induce condensation of $\alpha_s \bar{\alpha}_s$ for $s = 1, 2, \cdots n$ on the edge.
Since the different paired anyons $\alpha_s \bar{\alpha}_s$ and $\beta_{s'}\bar{\beta}_{s'}$ are mutual bosons, different incoherent channels do not compete with each other.
Thus, the number of possible phases characterized by anyon condensates is independent of the replica index $n$.
Hence, we expect that each phase of the EFD (equiv. $\tr\rho^2$) in Table~I has a correspondence in the replica limit $n \to 1$.

\bibliography{refs_short.bib}